\documentclass{ajour}

\usepackage{epsf}
\usepackage{cite}

\begin{document}

%


\authorrunninghead{Karyn Le Hur}

\titlerunninghead{Entanglement entropy of a dissipative two-level system}





\title{Entanglement entropy, decoherence, and quantum phase transitions of a dissipative two-level system}

\author{Karyn Le Hur}

\affil{Department of Physics, Yale University, New Haven, CT 06520, USA}

\email{karyn.lehur@yale.edu}

\abstract{The concept of entanglement entropy appears in multiple contexts, from black hole physics to quantum information theory, where it measures the entanglement of quantum states. We investigate the entanglement entropy in a simple model, the spin-boson model, which describes a qubit (two-level system) interacting with a collection of harmonic oscillators that models the environment responsible for decoherence and dissipation. The entanglement entropy allows to make a precise unification between entanglement of the spin with its environment, decoherence, and quantum phase transitions. We derive exact analytical results which are confirmed by Numerical Renormalization Group arguments both for an ohmic and a subohmic bosonic bath. Those demonstrate that the entanglement entropy obeys universal scalings. We make comparisons with entanglement properties in the quantum Ising model and in the Dicke model. We also emphasize the possibility of measuring this entanglement entropy using charge qubits subject to electromagnetic noise; such measurements would provide an empirical proof of the existence of entanglement entropy.}

\keywords{Dissipative two-level systems; Spin-boson models; Entanglement entropy; Decoherence; Quantum Phase Transitions}

\begin{article}

\section{Introduction}

Dissipation in quantum mechanics represents an important statistical mechanical problem having ramifications in systems as diverse as biological systems to limitations of quantum
computation. A prototype model in this class is the Caldeira-Leggett model describing a quantum particle in a dissipative bath (environment) of harmonic oscillators  \cite{caldeira}. The spin-boson model can be seen as a variant of the Caldeira-Leggett model where the quantum system is a two-level system  \cite{Blume,Leggett}. Those impurity systems are interesting because they display both a localized (classical) and delocalized (quantum) phase for the spin  \cite{karyn,karyn2}. This class of models were intensively investigated to study the emergent quantum-classical transition  \cite{Matthias}. 

An environment generally induces some ``uncertainty'' in the spin direction which is responsible for quantum decoherence, {\it i.e.}, for the rapid vanishing of the off-diagonal elements of the spin reduced density matrix. On the other hand, a finite coupling between the spin and its environment also induces entanglement between the spin and the environmental bosons, {\it i.e.}, the wavefunction of the system cannot be written as a simple product state anymore. In this paper, we are interested in quantifying properly the entanglement properties between the spin and its dissipative environment. One good measure of entanglement for these well-defined bipartite systems is the von Neumann entanglement entropy  \cite{entropy,Bennett,Amico}. In fact, we anticipate a deep connection between decoherence of the spin and (entanglement) entropy which may be seen as follows: the decoherence irreversibly converts the averaged or environmentally traced over spin density matrix from a pure state to a reduced mixture. Combining exact analytical results and Numerical Renormalization Group techniques applied directly on the spin-boson model \cite{karyn2,Bulla}, we are prompted to flesh out the entanglement properties between the spin and its environment by computing the von Neumann entropy of the spin when integrating over the bath degrees of freedom. We expand our recent articles  \cite{karyn3,karyn4}. We will demonstrate that the resulting entanglement entropy constitutes an important order parameter (which exhibits universal scalings), not only of quantum phase transitions but also of crossovers.  We will also confirm that loss of coherence of the two-level system (characterized either by the complete destruction of Rabi oscillations  \cite{Saleur} or by the strong suppression of persistent current in a ring  \cite{Buttiker,karyn3}) means a prominent enhancement in the entanglement of the system; this enhancement of entanglement corresponds to the actual quantum phase transition in the case of a subohmic bath and rather corresponds to the dynamical incoherent crossover from damped oscillatory to overdamped behavior in the case of an ohmic bath \cite{Leggett}. For a subohmic bath, second order phase transitions can be interpreted as transitions where coherence is lost due to the emergence of a purely classical state at the transition. We also emphasize the possibility of measuring the entanglement entropy of the spin using charge qubits subject to tunable electromagnetic noise  \cite{Schon,Clarke}. 

\section{Spin-Boson model, its relatives, Entropy}

In this section, we provide useful definitions. In particular, we introduce the spin-boson Hamiltonian as well as useful mappings towards long-range Ising chains  \cite{Dyson} and Kondo models  \cite{Hewson}, and we discuss more thoroughly the concept of entanglement entropy. We define properly the entanglement entropy $E$ of the spin with its environment in the context of the spin-boson model, and we discuss the different behaviors of $E$ in the delocalized and localized phase of the spin-boson model(s). The von Neumann entropy $E$ is strictly defined for a pure state at zero temperature; nevertheless, in Sec. 4.3 we will briefly comment on the quantum-classical crossover at finite temperature where the Boltzmann-Gibbs entropy proliferates \cite{Lebowitz}. 

\subsection{Spin-Boson model, Kondo physics, and Ising chains}

The Hamiltonian for the spin-boson model with a level asymmetry $h$ takes the usual form  \cite{Leggett}:
\begin{equation}
H_{SB}=-\frac{\Delta}{2}\sigma_x+\frac{h}{2}\sigma_z+H_{osc}+\frac{1}{2}\sigma_z \sum_n \lambda_n(a_n+a_n^{\dagger}),
\label{Hsb}
\end{equation}
where $\sigma_x$ and $\sigma_z$ are Pauli matrices and $\Delta$ is the tunneling amplitude between the states with $\sigma_z=\pm 1$.  $H_{osc}$ is the Hamiltonian of an infinite number of harmonic oscillators with frequencies $\{ \omega_n \}$, which couple to the spin degree of freedom via the coupling constants $\{ \lambda_n \}$.  The heat bath is characterized by its spectral function:
\begin{equation}
\label{Jw}
J(\omega) \equiv \pi \sum_n \lambda_n^2 \delta (\omega_n-\omega) =2\pi \alpha \omega_c^{1-s} \omega^s,
\end{equation}
where $\omega_c$ is a cutoff energy\footnote{To simplify notations, hereafter we fix the Planck constant $\hbar=1$.} and the dimensionless parameter $\alpha$ measures the strength of the dissipation. Similar models involving a two-level system coupled to a bath of harmonic oscillators are investigated in the context of stimulated photon emission. A known example is the Jaynes-Cummings model \cite{Jaynes}. It should be noticed that our model has a dense spectrum of fundamental frequencies of the harmonic oscillators.

In the following, we will refer to the values $0<s<1$ as subohmic damping \cite{Leggett,Matthias}. One standard approach is to integrate out the dissipative bath, leading to an effective action which is reminiscent of the classical spin chains with long-range correlations (here, in time):
\begin{equation}
{\cal S}_{int} = \int d\tau d\tau' \sigma_z(\tau) g(\tau-\tau') \sigma_z(\tau'),
\end{equation}
with 
\begin{equation}
g(\tau)\propto 1/\tau^{1+s},
\end{equation}
at long times. 
\begin{figure}[!ht]
\begin{center}
\epsfxsize=1\textwidth{\epsfbox{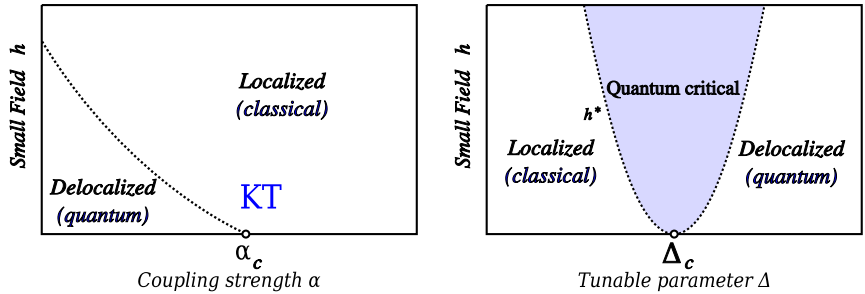}}
\end{center}
\caption{\label{hasymmetry} Phase diagram of the spin-boson model for the ohmic and subohmic case.}
\end{figure}
For a subohmic bath, correlations in time become sufficiently long-range that one may expect the localization of the spin for all values of $\alpha>0$, at least in the limit of $\Delta\rightarrow 0$. On the other hand, by increasing $\Delta$, one eventually expects a quantum phase transition towards a delocalized phase for the spins. In fact, the analogy with classical spin chains leads to the existence of second-order quantum transitions for $0<s<1$ between a localized phase at small $\Delta$ and a delocalized phase at large $\Delta$  \cite{Dyson,Kosterlitz}; see the phase diagram of Fig. 1. It is relevant to mention that this transition has been seen by applying the Numerical Renormalization Group (NRG) method directly on the spin-boson model \cite{Bulla,karyn4,Matthias} but this is not visible in the popular non-interacting blip approximation \cite{Leggett}.   

In the localized regime, the tunnel splitting $\Delta$ between the two levels renormalizes to zero, {\it i.e.}, the spin gets trapped in one of the states $\uparrow$ or $\downarrow$, whereas $\Delta$ is strongly renormalized in the delocalized phase. 

The special case of $s=1$ corresponds to the ohmic spin-boson model which is also equivalent to the anisotropic Kondo model; at small $\Delta$, here the system shows a Kosterlitz-Thouless (KT) type quantum phase transition\footnote{In the ohmic case, the phase transition is described by RG equations similar to those in the XY model in two dimensions. On the other hand, the phase transition in the Kondo model was found earlier than the KT phase transition \cite{anderson}.}, separating the localized phase at $\alpha\geq \alpha_c$ from the delocalized phase at $\alpha<\alpha_c$. To understand how the KT transition takes place as a function of the dissipation parameter $\alpha$, one may
use the analogy with the anisotropic Kondo model, which describes the exchange interaction between conduction electrons and a magnetic impurity \cite{anderson}:
\begin{eqnarray}
H_{K}&=&H_{\mbox{\scriptsize kin}}+\frac{J_{\perp}}{2} \sum_{kk^{\prime}} \left( c_{k\uparrow}^{\dagger}c_{k^{\prime}\downarrow} S^- + c_{k\downarrow}^{\dagger}c_{k^{\prime}\uparrow} S^+ \right) \nonumber \\
& & +\frac{J_z}{2}\sum_{kk^{\prime}} \left( c_{k\uparrow}^{\dagger}c_{k^{\prime}\uparrow}-c_{k\downarrow}^{\dagger}c_{k^{\prime}\downarrow} \right) S_z+hS_z,
\end{eqnarray}
where $H_{\mbox{\scriptsize kin}}$ represents the kinetic energy of the electrons. The equivalence between the spin-boson model with ohmic dissipation and the Kondo model has been first realized by Chakravarty  \cite{chakravarty} and independently by Bray and Moore  \cite{bray}. One may identify:
\begin{equation}
(\rho J_{\perp}) \longrightarrow \frac{\Delta}{\omega_c}
\hskip 0.2cm \hbox{and} \hskip 0.2cm (1+2\delta_e/\pi)^2 \longrightarrow \alpha,
\end{equation}
where $\rho$ is the conduction electron density of states and the parameter $\delta_e$  is related to the
phase shift caused by the $J_{z}$ Kondo term and is given by \cite{Guinea}
\begin{equation}
\delta_e = \tan^{-1}(-\pi \rho J_z/4).
\end{equation}
The phase transition in the Kondo model (a purely quantum phase transition at the absolute zero) corresponds to the (bare) Kondo coupling changing sign. For $J_z>0$ ({\it i.e.}, antiferromagnetic coupling of the Kondo spin to conduction electrons), the ``fugacity'' $J_{\perp}$ is a relevant operator. Physically, this means that the number of spin flips proliferates, and the Kondo spin forms a singlet state with the conduction electrons. In the opposite $J_z<0$ (ferromagnetic Kondo coupling) case, $J_{\perp}$ scales to zero. No spin flips remain, and the spin is ``frozen'' in time corresponding to the long range order of the Ising chain. Anderson, Yuval, and Hamman \cite{anderson} succeeded in mapping the behavior of the Kondo model onto the solution of a long-range Ising spin chain. The tunneling events where the local moment spin flips due to scattering with conduction electrons, correspond to domain walls of the equivalent Ising chain. The problem is similar to ordinary tunneling of a two-level system. The physically new feature
is the fact that conduction electrons represent a dissipative bath, and as result the tunneling events are not independent. Instead, they feature long-range interactions in time, and the equivalent Ising chain
acquires long-range interactions. 

The equivalence between the anisotropic Kondo model and the spin-boson model with ohmic damping
can be formulated explicitly through bosonization  \cite{Guinea}. The delocalized region corresponds to the antiferromagnetic Kondo model, while the localized region corresponds to the ferromagnetic Kondo model where the spin is frozen in time. In the delocalized phase of the spin-boson model, the Kondo energy scale obeys\footnote{In the definition of the Kondo energy scale, we have introduced a high-energy cutoff $D$ for the conduction electrons which is of the order of the Fermi energy; the relation between the two cutoffs $\omega_c$ and $D$ will be fixed properly later; see Appendix C.} \cite{Bohr}
\begin{equation}
T_K=\Delta (\Delta/D)^{\alpha/(1-\alpha)},
\end{equation}
 for values of $\alpha$ not too close to the transition and close to the transition $T_K$ assumes the exponential form of the isotropic, antiferromagnetic Kondo model; $\ln T_K \propto 1/(\alpha_c-\alpha)$.
The critical line separating the localized and delocalized phase in the spin-boson model thus corresponds to $\alpha_c = 1 +{\cal O}(\Delta/\omega_c)$ or $J_z=-|J_{\perp}|$.
 Such results have been confirmed by applying the NRG approach directly to the spin-boson model \cite{karyn2,Bulla}. In the delocalized phase of the spin-boson model, 
we expect a unique nondegenerate ground state and the expansion in the number of spin flips  in the partition function does not lead to a dilute gas of flips for $T\ll T_K$ which reflects the basic difficulty that the exact ground state is orthogonal to that for a static spin $\Delta=0$; consult Sec. 3.1. 

\subsection{Entanglement entropy}

The problem of measuring entanglement in many-body systems is a lively field of research. Bipartite
entanglement of pure states is conceptually well understood  \cite{Amico,Gil,Jordan}.  A useful measure of many-body entanglement when the total system is in a pure state is the von Neumann entanglement entropy  \cite{entropy, Bennett, Amico}. This is obtained by focusing on bipartite systems where space can be divided into 2 regions, A and B. Beginning with the ground state pure density matrix, region B is traced over to define the reduced density matrix $\rho_A$. From this, the von Neumann entanglement entropy 
\begin{equation}
E(r) = -\hbox{Tr}[\rho_A \log_2 \rho_A], 
\end{equation}
is obtained for the subsystem A of size $r$. The rate at which $E$ grows  with the spatial extent, r, of region $A$ is not only a fundamental measure of entanglement, it is also crucial to the functioning of the Density Matrix Renormalization Group \cite{DMRG}. For systems with finite correlation lengths, it is generally expected that $E$ grows with the area of the boundary of region A  \cite{bombelli}. In the one-dimensional case, conformally invariant systems have $E(r) \rightarrow (c/3) \ln r$ where c is the ``central charge'' characterizing the conformal field theory \cite{c,c2,Joel}; for a two-dimensional example, see Ref. \cite{Fradkin}. Entanglement entropy has recently been shown to be a useful way  of characterizing topological phases of many body theories \cite{Kitaev,Fendley,Wen}, which cannot 
be characterized by any standard order parameter. Entanglement entropy is also closely 
related to the thermodynamic entropy of black holes and to the ``holographic principle''
relating bulk to boundary field and string theories  \cite{Fendley,Ryu}. 
Entanglement entropy also brings a new light on quantum impurity systems; a spin can be entangled to a bath of conduction electrons  \cite{Affleck} or to a dissipative bosonic environment  \cite{Angela,Costi,karyn3,karyn4}. For a quantum impurity system (A is the spin):
\begin{equation}
E=-p_+ \log_2 p_+ - p_- \log_2 p_- , 
\end{equation}
where 
\begin{equation}
p_{\pm}=\left( 1\pm \sqrt{\langle \sigma_x \rangle^2 + \langle \sigma_y \rangle^2+ \langle \sigma_z \rangle^2}\right)/2.
 \end{equation}
 For the spin-boson model, $\langle \sigma_y \rangle =0$ because $H_{SB}$ is invariant under 
 $\sigma_y \to -\sigma_y$.  Moreover, for the spin-boson model, we get the equalities:
 \begin{eqnarray}
 \langle \sigma_x \rangle &=& -2\partial {\cal E}_g/\partial \Delta \\
 \langle \sigma_z \rangle &=& 2\partial {\cal E}_g/\partial h,
 \end{eqnarray}
where ${\cal E}_g$ is the ground state energy of $H_{SB}$. Since $H_{SB}$ and $H_{AKM}$ are related by a unitary transformation, they have the same ground state energy (up to an unimportant constant).  The field $h$ couples directly to the spin in both models, so we have $\langle S_z \rangle=\langle \sigma_z \rangle/2$.  A similar relationship does {\em not} hold between $\langle S_x \rangle$ and $\langle \sigma_x \rangle$, and thus $E$ does not (exactly) measure the entanglement between the Kondo impurity and the conduction band. 
 
 At $\alpha=0$, the qubit is decoupled from the environment; thus, $\langle \sigma_x \rangle=\Delta/\sqrt{h^2+\Delta^2}$ and $\langle \sigma_z \rangle=-h/\sqrt{h^2+\Delta^2}$.  With $p_+=1$ and $p_-=0$, we check that $E=0$ for all values of $\Delta$ and $h$. Moreover, at large $h$, we also must have $E \to 0$  because the qubit is localized in the state with $\langle \sigma_z \rangle = -1$ and $\langle \sigma_x \rangle = 0$. In a similar way, in the localized phase, since the qubit is also localized in one
 classical state at arbitrarily small $h$ then $E\rightarrow 0$\footnote{This result is intuitive if one absorbs the coupling with the environment into the tunneling term \cite{Buttiker}. In the localized phase, the tunneling term is renormalized (almost) to zero which then ensures a disentanglement (decoupling) with the environment.}.
 Below, we want to understand how $E$ interpolates between all these limits. We argue that the entanglement entropy of the spin with its environment will allow us to make a unification between entanglement of the spin with its environment, decoherence, and quantum phase transitions. Entanglement near a quantum
 phase transition is of great current interest \cite{Amico,Osborne,Osterloh}.
 
 It is also relevant to note the relation between entanglement and quantum decoherence of the spin.
 Quantum decoherence implies a rapid reduction of the off-diagonal elements of the spin density matrix,
 {\it i.e.}, $\langle \sigma_x\rangle=0$ which results in $p_{\pm}\rightarrow 1/2$ and $E\rightarrow 1$ if $\langle \sigma_z\rangle$ vanishes at small $h$ (delocalized phase).  We will show below that for the ohmic case, the quantum decoherence or maximal entanglement occurs at the Toulouse limit 
 $\alpha=1/2$ \cite{Toulouse} whereas for the subohmic case this rather happens at the phase transition.
 
\section{Ohmic case: Spin Observables and Entanglement}

After these generalities, we consider the ohmic situation and a large level asymmetry $\Delta/h\ll 1$ where perturbation theory applies and a correspondence to the well-known $P(E)$ theory for dissipative systems can be formulated \cite{Nazarov}. We are also interested in extracting the spin observables.

\subsection{Large level asymmetry and P(E) theory}

\begin{figure}[!ht]
\begin{center}
\epsfxsize=0.35\textwidth{\epsfbox{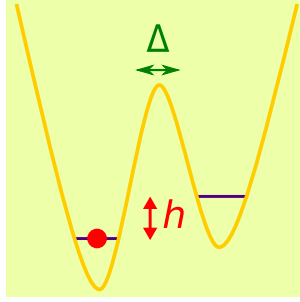}}
\end{center}
\caption{\label{hasymmetry} Two-level system with a large level asymmetry $\Delta/h\ll 1$.}
\end{figure}

For $\Delta=0$, the spin is completely localized in the $\downarrow$ state. Now, for finite $\Delta$, we want to evaluate the probability $p_{del}$ that this spin down electron flips up. In the absence of the environment, it is straightforward to obtain that $p_{del} = \Delta^2/h^2 +...$ . Below, we will incude the effect of the environment and see how this quantity gets modified. It should be noted that $p_{del}$ plays an important role because it is related to $\langle \sigma_z\rangle$, through 
\begin{equation}
\langle \sigma_z\rangle +1\sim {\cal O}(p_{del}). 
\end{equation}

For $\Delta=0$, one gets two classical states and the Hamiltonian for either of the states depends on the bosons:
\begin{equation}
H_{\uparrow,\downarrow} = \sum_n \omega_n a^{\dagger}_n a_n \pm \sum_n \frac{\lambda_n}{2}(a_n+a_n^{\dagger}) +cst.
\end{equation}
We can easily handle this Hamiltonian when $\lambda_n=0$. The stationary wavefunctions are those
with a fixed number of bosons $n_n=a^{\dagger}_n a_n$ in each mode corresponding to the energy $\sum_n \omega_n n_n$. For $\lambda_n \neq 0$, one may absorb the linear term in the redefinition 
\begin{eqnarray}
a_{n,\uparrow} &=& a_n + \frac{\gamma_n}{2} \\ \nonumber
a_{n,\downarrow} &=& a_n - \frac{\gamma_n}{2}.
\end{eqnarray}
where $\gamma_n = \lambda_n/\omega_n$.
Since the bosons have been shifted, the vacua for the two states are not the same. More precisely, let us consider the vacuum
\begin{equation}
a_{m,\downarrow}|O_{\downarrow}\rangle =0.
\end{equation}
This is equivalent to 
\begin{equation}
a_{m,\uparrow}|O_{\uparrow}\rangle = \gamma_m |O_{\uparrow}\rangle,
\end{equation}
We then observe that $|O_{\downarrow}\rangle$ is not the vaccum for the bosons linked to the state $\uparrow$. This is rather a coherent state and thus
\begin{equation}
|O_{\downarrow}\rangle = \exp\left(-\frac{|\gamma_m|^2}{2}\right)\sum_{n_m=0}^{+\infty} \frac{(\gamma_m)^{n_m}}{\sqrt{n_m!}}|\{n_m\}_{\uparrow}\rangle.
\end{equation}
For convenience of notations, in this expression we have assumed that $m$ is fixed.
To first order in $\Delta$, this implies that the ground state
$|O_{\downarrow}\rangle$ acquires corrections proportional to all possible states $|\{n_m\}_{\uparrow}\rangle$:
\begin{equation}
|g\rangle = |O_{\downarrow}\rangle + \sum_{\{n_m\}} \psi\{n_m\}|\{n_m\}_{\uparrow}\rangle,
\end{equation}
with
\begin{equation}
\psi\{n_m\} = \Delta \frac{\langle O_{\downarrow}|\{n_m\}_{\uparrow}\rangle}{h
+ n_m \omega_m}.
\end{equation}
The probability that the spin flips up, then takes the form:
 \begin{equation}
 p_{del} = \sum_{\{n_m\}} \left|\psi{\{n_m\}}\right|^2+...\  .
 \end{equation} 
 
By carefully summing over all modes $m$, one can make a formal link with the $P(E)$ theory \cite{Nazarov} of dissipative tunneling problems (see Appendix A):
\begin{equation}
p_{del} = \Delta^2 \int_0^{\omega_c} dE \frac{P(E)}{(h+E)^2}+...\  .
\end{equation}
If there is no dissipation, then $P(E)=\delta(E)$, and thus this reduces to the known qubit result
$p_{del} = \Delta^2/h^2+...$ . In the case of Nyquist noise or ohmic dissipation where $P(E)\propto E^{2\alpha-1}$ \cite{Nazarov,karyn5}, the delocalization probability rather evolves as
\begin{equation}
p_{del} = \frac{\Delta^2}{h^2}\left(\frac{h}{\omega_c}\right)^{2\alpha}.
\end{equation}
Note that $p_{del}$ is small at high $h \sim \omega_c$
and one reproduces $p_{del} \sim \Delta^2/h^2$. On the other hand, for $\alpha<1$, one observes that $p_{del}$ increases for smaller $h$ and eventually reaches the maximum value $p_{del}\rightarrow 1$. This shows that the problem becomes highly non-perturbative at small $h$  for $\alpha<1$; one indeed expects a delocalized phase for $\alpha<1$ and a localized phase for $\alpha>1$. {\it It is important to
note that for small level asymmetries,  the point $\alpha=0$ is unstable, {\it i.e.}, perturbation theory in 
$\Delta$ already breaks down at small $\alpha$}.

The result in Eq. (24) is in agreement with the expansion of the ground state energy to second order in $\Delta$ (we consider the case where $h>0$) \cite{Buttiker}:
\begin{equation}
\label{pert}
{\cal E}_g = -\frac{h}{2} -\frac{\omega_c}{4}\left(\frac{\Delta}{\omega_c}\right)^2 e^{h/\omega_c}\left(\frac{h}{\omega_c}\right)^{2\alpha-1} \Gamma\left(1-2\alpha,\frac{h}{\omega_c}\right),
\end{equation}
where $\Gamma$ is the incomplete gamma function. In the regime of relatively large $h$ (but still
small compared to the high-energy cutoff $\omega_c$), one may simplify
\begin{equation}
\Gamma \left(1-2\alpha,\frac{h}{\omega_c}\right) \simeq \Gamma(1-2\alpha) - \frac{\left(h/\omega_c\right)^{1-2\alpha}}{1-2\alpha}.
\end{equation}
This immediately leads to:
\begin{eqnarray}
{\cal E}_g &=& -\frac{h}{2} +\frac{\omega_c}{4}\left(\frac{\Delta}{\omega_c}\right)^2 
e^{h/\omega_c}\\ \nonumber
&\times& \left[\frac{1}{1-2\alpha} - \Gamma(1-2\alpha)\left(\frac{h}{\omega_c}\right)^{2\alpha-1}\right].
\end{eqnarray}
Thus, $\langle \sigma_z\rangle = 2\partial {\cal E}_g/\partial h$ can easily be derived. At relatively large values of the level asymmetry, one gets:
\begin{equation}
\langle \sigma_z\rangle = -1 +\frac{1}{2}\left(\frac{\Delta}{\omega_c}\right)^2(1-2\alpha) \Gamma(1-2\alpha)\left(\frac{h}{\omega_c}\right)^{2\alpha-2}.
\end{equation}
We check that $\langle \sigma_z\rangle +1\sim {\cal O}(p_{del})$. In a similar way, one can extract
$\langle \sigma_x\rangle = -2{\cal E}_g/\partial \Delta$, resulting in:
\begin{equation}
\label{pertsx}
\langle \sigma_x\rangle = -\frac{\Delta}{\omega_c}\left[\frac{1}{1-2\alpha} - \Gamma(1-2\alpha)\left(\frac{h}{\omega_c}\right)^{2\alpha -1}\right].
\end{equation}

It should be noted that the limit $\alpha\rightarrow 1/2$ is always well-defined, since
\begin{equation}
\lim_{x\rightarrow 0}\left(\frac{1}{x}-\Gamma(x)y^{-x}\right) = \lim_{x\rightarrow 0} \frac{1-y^{-x}}{x} = \ln y.
\end{equation}
In particular, this leads to:
\begin{eqnarray}
\langle \sigma_x\rangle_{\alpha=1/2} &=& -\frac{\Delta}{\omega_c}\ln\left(\frac{h}{\omega_c}\right),\\ \nonumber
\langle \sigma_z\rangle_{\alpha=1/2} &=& -1 +\frac{\Delta^2}{2\omega_c h},
\end{eqnarray}
in agreement with the exact results obtained from the non-interacting resonant level model at the Toulouse point $\alpha=1/2$ \cite{Toulouse}; see Appendix B. This implies the important equality between the two cutoffs of the theory:
\begin{equation}
D(\alpha=1/2) = \frac{4\omega_c}{\pi}.
\end{equation}

\section{Results for a vanishing level asymmetry}

In the delocalized phase, one can use the Bethe Ansatz solution of the anisotropic Kondo model or of the equivalent interacting resonant level model; consult Ref. \cite{Level,ponomarenko} and Appendix C. While general expressions are quite complicated \cite{Buttiker}, it is instructive to derive simple {\it  scaling} forms for the observables in the limits  $h \ll T_K$ and $h \gg T_K$ \cite{karyn3}. We will also apply the NRG method which shows an excellent agreement with the Bethe Ansatz results; NRG parameters have been defined in our Ref. \cite{karyn4}.

\subsection{Spin observables}

 For $h \gg T_K$, we check that the Bethe Ansatz calculations reproduce the perturbative results of Sec. 3.1. For the sake of clarity, technical details are given (hidden) in Appendix C.
For $h \ll T_K$, we obtain,
\begin{equation}
\lim_{h \ll T_K} \langle \sigma_z \rangle = -\frac{2e^{\frac{b}{2(1-\alpha)}}}{\sqrt{\pi}}  \frac{\Gamma[1+1/(2-2\alpha)]}{\Gamma[1+\alpha/(2-2\alpha)]} \left( \frac{h}{T_K} \right),
\end{equation}
where 
\begin{equation}
b=\alpha \ln \alpha + (1-\alpha) \ln (1-\alpha). 
\end{equation}
Note that $\langle \sigma_z \rangle \propto h/T_K$ at small $h$, in keeping with the Kondo Fermi liquid ground state. The local
susceptibility of the spin converges to $1/\Delta$ for $\alpha\rightarrow 0$\footnote{For $\alpha\rightarrow 0$, $\Gamma[1]=1$, $\Gamma[3/2]=\sqrt{\pi}/2$, $\exp(b/(2(1-\alpha))=1$, and $T_K=\Delta$.}  in accordance with the
two-level description and diverges in the vicinity of the KT phase transition as a result of the exponential suppression of the Kondo energy scale.  {\it It is relevant to observe that the longitudinal spin magnetization $\langle \sigma_z\rangle$ only depends on the ``fixed point'' properties, {\it i.e}, this is a universal function of $h/T_K$ in the delocalized phase.} Exactly at the KT transition,
one can adapt the results of Anderson and Yuval  \cite{Anderson2} and the result is that $\langle \sigma_z\rangle$ jumps by the amount $\sqrt{1/\alpha_c}$ along the quantum critical line where $\alpha_c = 1 +{\cal O}(\Delta/\omega_c)$. 
\begin{figure}[!ht]
\begin{center}
\epsfxsize=0.68\textwidth{\epsfbox{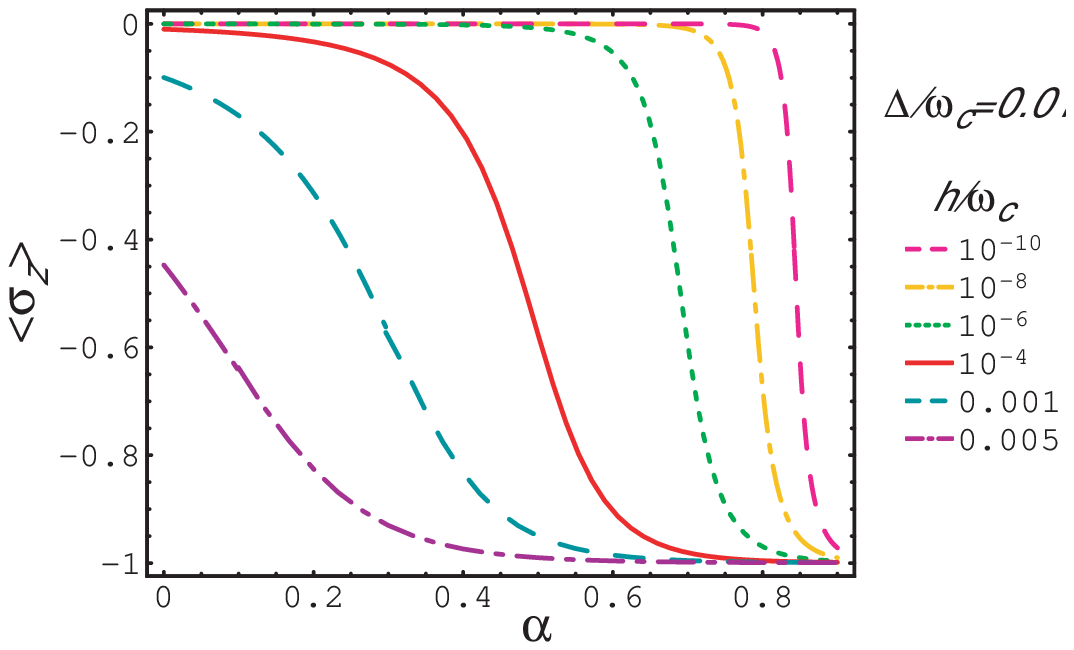}}
\epsfxsize=0.70\textwidth{\epsfbox{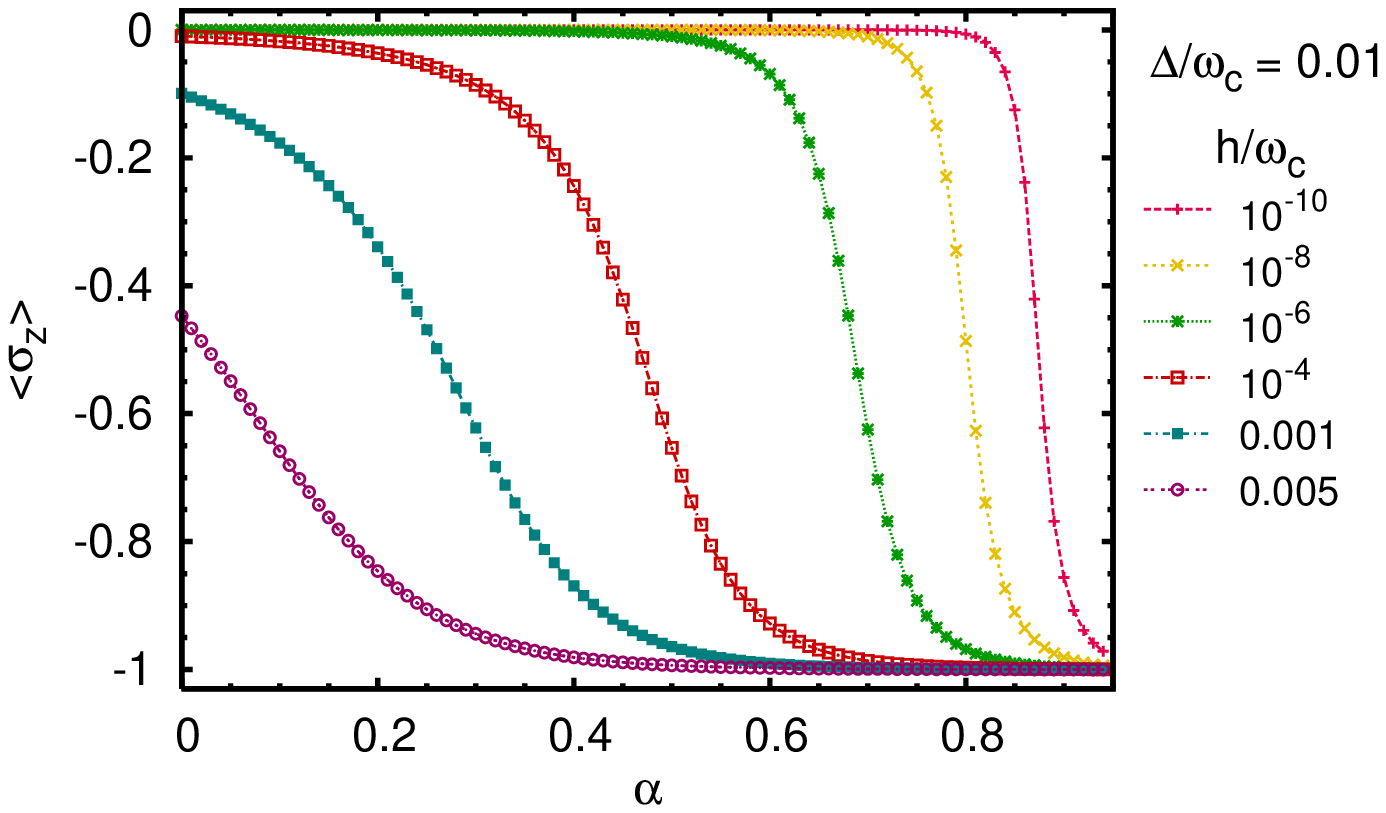}}
\end{center}
\caption{\label{Sz} Longitudinal spin magnetization $\langle \sigma_z\rangle$ as a function of the detuning $h$ from Bethe Ansatz techniques (top) and NRG calculations (bottom).}
\end{figure}
This results in a {\it non-universal} jump $\sim 1-{\cal O}(\Delta/\omega_c)$ at the KT transition \cite{karyn2}. Remember that in contrast to the universal jump of the superfluid density in two-dimensional XY models, in the spin-boson model, the jump in the longitudinal magnetization is non-universal for finite $\Delta/\omega_c$. Finally, far in the localized phase, one rather predicts $\langle \sigma_z\rangle \approx -1 +{\cal O}((\Delta/\omega_c)^2)$ \cite{Angela}.
In Fig. \ref{Sz} we present two curves of $\langle \sigma_z\rangle$ as a function of $h$, which have been obtained with the Bethe Ansatz and with the NRG calculations, respectively. One can notice the very good agreement between the two approaches. By increasing the detuning $h$, the abrupt jump in $\langle \sigma_z\rangle$ occurring at the quantum phase transition is progressively replaced by a smooth behavior; 
at finite $h$ there is a smooth crossover separating the delocalized and the localized regime. 

The leading behavior of $\langle \sigma_x \rangle$ in the delocalized phase takes the form:
\begin{equation}
 \lim_{h \ll T_K} \langle \sigma_x\rangle = \frac{1}{2\alpha-1} \frac{\Delta}{\omega_c}+C_1(\alpha) \frac{T_K}{\Delta},
 \label{sigxh0}
 \end{equation}
 \begin{figure}[!ht]
\begin{center}
\epsfxsize=0.52\textwidth{\epsfbox{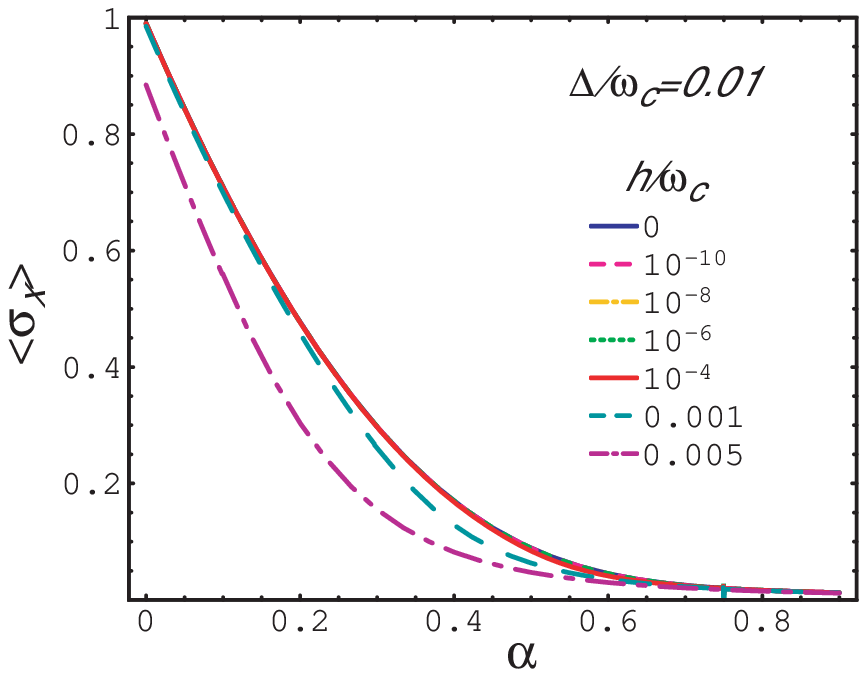}}
\epsfxsize=0.57\textwidth{\epsfbox{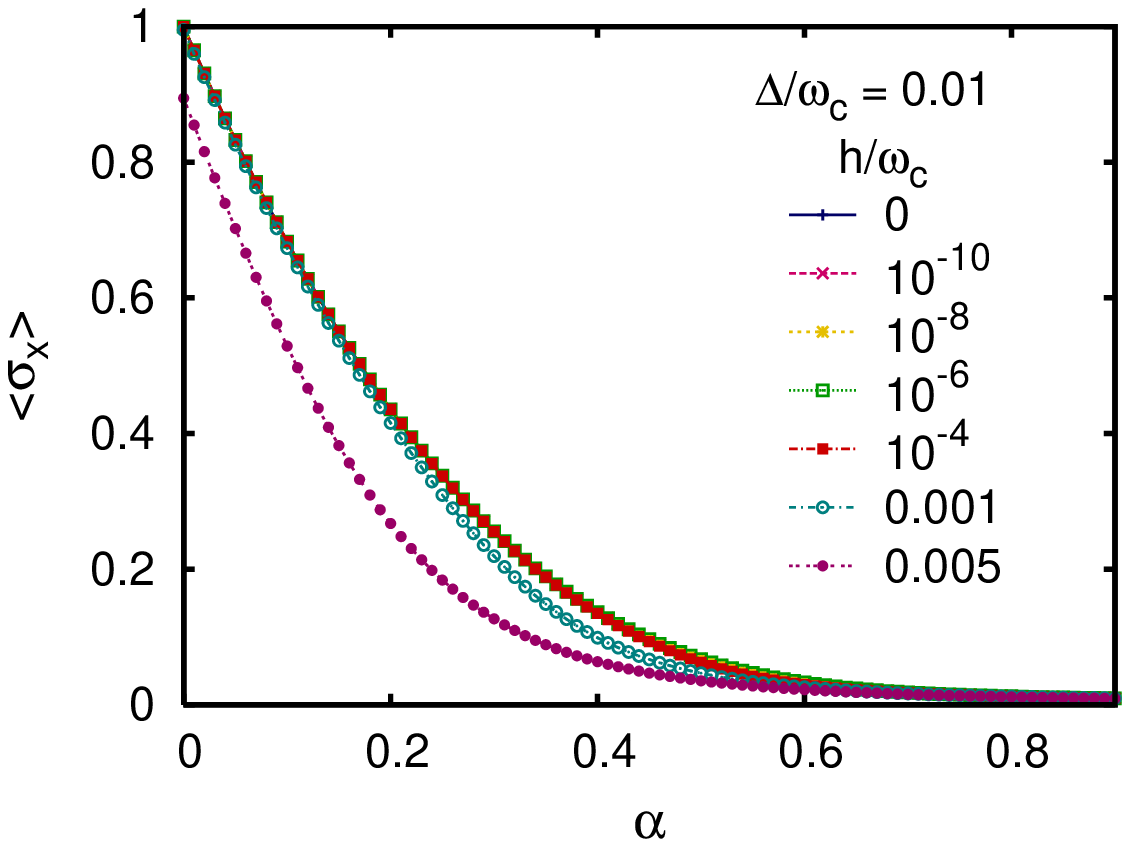}}
\end{center}
\caption{\label{Sx} The transverse spin magnetization $\langle \sigma_x \rangle$ versus $\alpha$ for different level
asymmetries $h$ obtained from the Bethe Ansatz calculations (top) and from NRG (bottom).}
\end{figure}
with 
\begin{equation}
C_1(\alpha)  =  \frac{e^{-b/(2-2\alpha)}}{\sqrt{\pi}(1-\alpha)} \frac{\Gamma[1-1/(2-2\alpha)]}{\Gamma[1-\alpha/(2-2\alpha)]}.
\end{equation}
As $\alpha \to 0$, $T_K \to \Delta$ and $C_1(0)=1$, so we recover the exact result $\langle \sigma_x \rangle_{\alpha=h=0} =1$\footnote{Exactly at $\alpha=0$, the (small) first term in Eq.  (\ref{sigxh0}) is not present; see Appendix C.}. As we turn on the coupling to the environment, we introduce some uncertainty in the spin direction and $\langle \sigma_x \rangle$ decreases. {\it It should be noted that $\langle \sigma_x\rangle$ does not only depend on fixed point properties; in the delocalized phase, $\langle \sigma_x\rangle$ still contains a perturbative part in $\Delta/\omega_c$!}

For $\alpha < 1/2$, the monotonic decrease of $T_K/\Delta$ dominates.  In contrast, for $\alpha>1/2$, the first term in Eq.~(\ref{sigxh0}) dominates and we have  
\begin{equation}
\langle \sigma_x\rangle_{\alpha>1/2,h\rightarrow 0} = \frac{1}{2\alpha-1}\frac{\Delta}{\omega_c}.
\end{equation}
In fact, this result can also be recovered using the perturbation theory of Sec. 3.1 (consult Eq. (\ref{pertsx}) for $h\rightarrow 0$). This result clearly emphasizes that the observable $\langle \sigma_x\rangle$ is {\it continuous} and {\it small} at the KT transition in the ohmic spin-boson model.  This is also consistent with the work by Anderson and Yuval which predicts $\langle \sigma_x\rangle \sim \Delta/\omega_c$ exactly at the phase transition \cite{Anderson2}. 

The fact that $\langle \sigma_x\rangle$ becomes very small $\sim \Delta/\omega_c$ in the localized phase can be understood from  the mapping between the anisotropic Kondo model and the Ising model with long-range correlations in time. Here, the partition function of the system can be expanded in powers of $\Delta^2$  \cite{Leggett,anderson,Anderson2}:
\begin{equation}
Z = \sum_n \frac{\bar{\Delta}^{2n}}{2n!} \int_0^{\beta} \frac{d\tau_1}{\tau_c}... \int_0^{\beta} \frac{d\tau_{2n}}{\tau_c} \prod_{ij=1,...,2n}
{\cal W}\left(\frac{\tau_i-\tau_j}{\tau_c}\right),
\end{equation}
where the function ${\cal W}$ describes the (long-range) interaction between kinks (spin flips) located at positions $\tau_i$ and $\tau_j$, and $\bar{\Delta}=\Delta/\omega_c$. 

The scaling procedure requires to modify the short-time cutoff $\tau_c=\omega_c^{-1}
\rightarrow \tau_c-d\tau_c$. Two important features then appear by modifying $\tau_c$: the dependence
of ${\cal W}\left(\frac{\tau_i-\tau_j}{\tau_c}\right)$ can be included in a global renormalization of $\Delta$
and configurations with a kink-antikink pair at distance between $\tau_c$ and $\tau_c-d\tau_c$ have to 
be replaced by configurations where this pair is absent.  The number of removed pairs is proportional to
$d\tau_c$ and the (free) energy variation is roughly $-\bar{\Delta}^2(\tau_c)\tau_c^{-2} d\tau_c$:
\begin{equation}
\frac{\partial F}{\partial \tau_c} = -\bar{\Delta}^2(\tau_c)\tau_c^{-2},
\end{equation}
This is equivalent to change $\omega_c\rightarrow \omega_c+d\Lambda=\Lambda$:
\begin{equation}
\frac{\partial F}{\partial \Lambda} = \bar{\Delta}^2(\Lambda).
\end{equation}
The renormalization of $\bar{\Delta}$ in the ohmic case is  \cite{Leggett}:
\begin{equation}
\bar{\Delta}(\Lambda) = \bar{\Delta} \left(\frac{\Lambda}{\omega_c}\right)^{\alpha-1}.
\end{equation}
This leads to
\begin{equation}
F(\Lambda) = \int_{\omega_c}^{\Lambda} d\Lambda \bar{\Delta}^2(\Lambda).
\end{equation}
The ground state energy can be recovered by taking the limit $\Lambda\rightarrow 0$. This is continuous at the quantum phase transition. Extending this result to the delocalized phase, we
find that for $\alpha\geq 1/2$ this expression is effectively convergent at low energies, and essentially
one recovers the result obtained by Bethe Ansatz or by perturbation theory:
\begin{equation}
{\cal E}_g(\Lambda=0) \approx - \frac{\omega_c}{(2\alpha-1)}\left(\frac{\Delta}{\omega_c}\right)^2.
\end{equation}
This ensures that in the delocalized phase, for $\alpha>1/2$, $\langle \sigma_x\rangle$ is of the order of $\Delta/\omega_c$. This result is definitely emphasized in Figs. \ref{Sx}. In fact, it should be noted that
$\langle \sigma_x\rangle$ is always continuous at a quantum phase transition.

Finally, we can also check that $\langle \sigma_x\rangle$ evolves continuously close to $\alpha=1/2$. 
In the limit $\alpha \to 1/2$, one can take $C_1(\alpha) = (4/\pi) \Gamma(1-2\alpha) \to 4/(\pi (1-2\alpha))$ and use the identity $D(\alpha=1/2)=4\omega_c/\pi$ to find $\langle \sigma_x \rangle \to -(4/\pi) \sqrt{T_K/D} \ln (T_K/D)$, in agreement with the ``non-interacting'' resonant level description valid at the very specific point $\alpha=1/2$.

\subsection{Entanglement entropy}

\begin{figure}[!ht]
\begin{center}
\epsfxsize=0.65\textwidth{\epsfbox{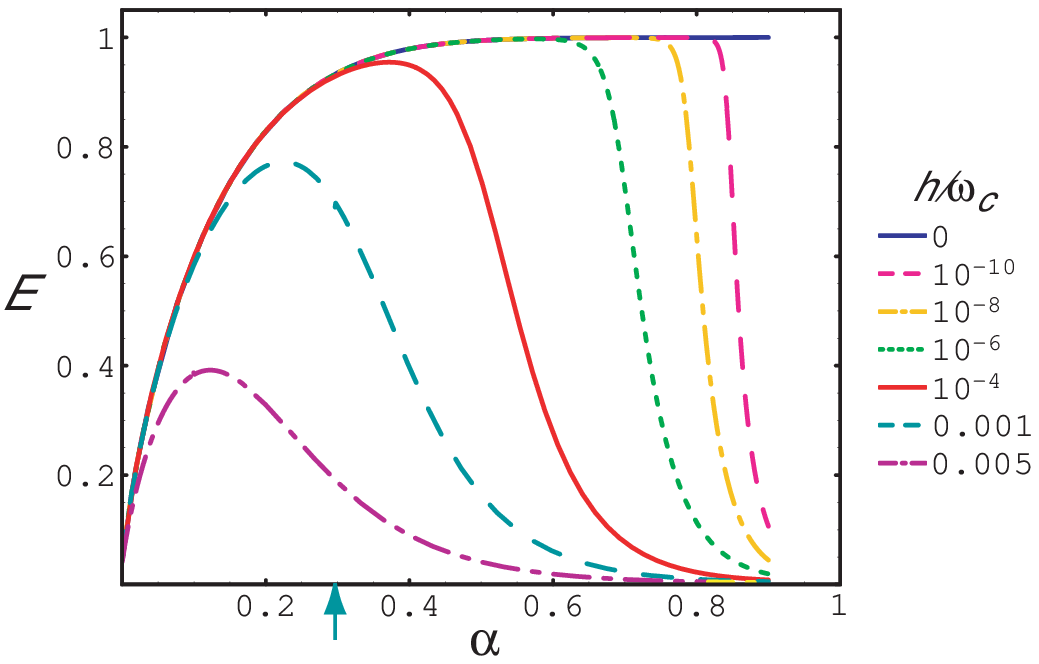}}
\epsfxsize=0.7\textwidth{\epsfbox{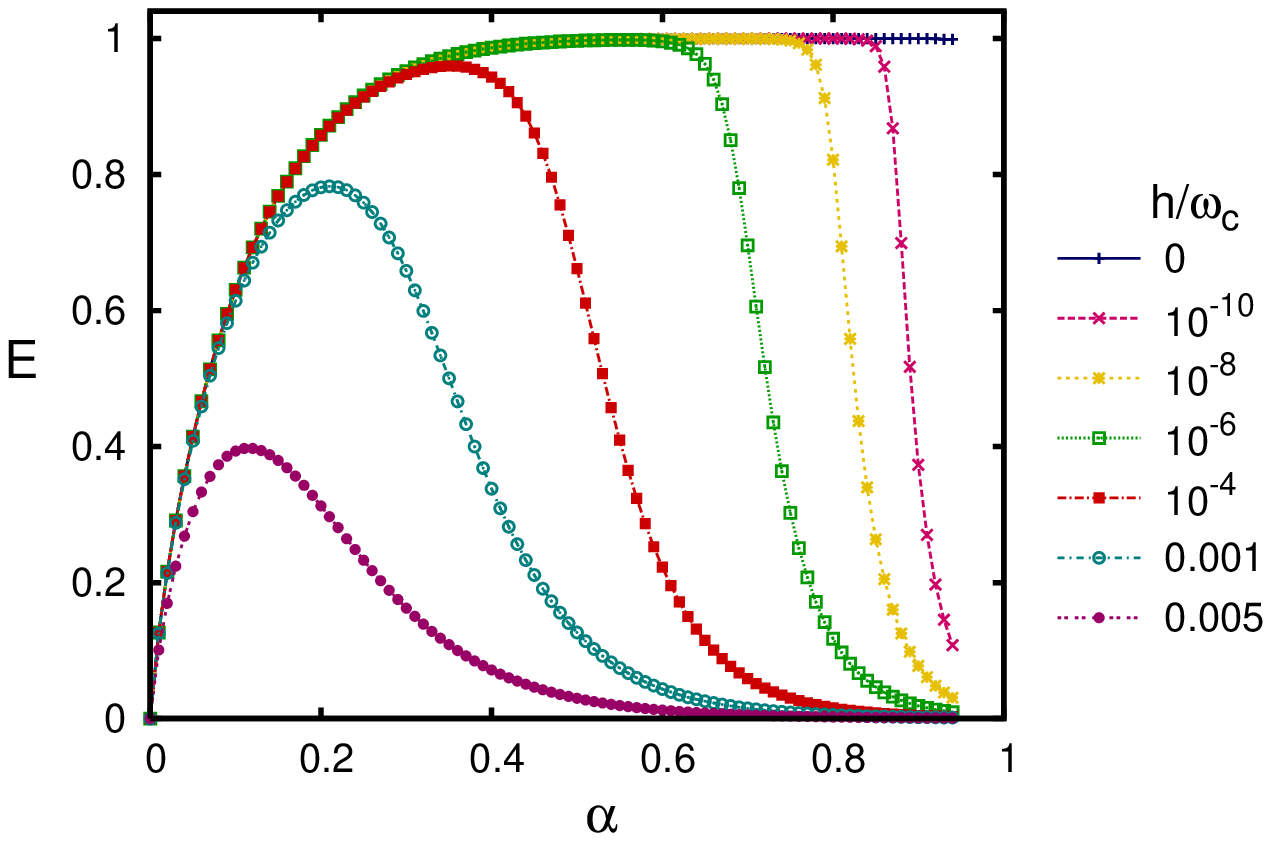}}
\end{center}
\caption{\label{entropy} $E(\alpha,\Delta=0.01 \omega_c,h)$ versus $\alpha$ at several values of $h$ from Bethe Ansatz calculations (top) and NRG. The NRG curve is in agreement with the one of Ref. \cite{Costi}.}
\end{figure}

Now, we are able to study thoroughly the entanglement between the spin and its environment, at least in the context of an ohmic bath; see Fig. \ref{entropy}. 

First, in the delocalized phase, at $h=0$, $E$ is entirely determined by $\langle \sigma_x\rangle$. Thus,
at small $\alpha$, $E$ grows linearly with $\alpha$ and for $\alpha\geq 1/2$, $\langle \sigma_x\rangle \approx \Delta/\omega_c \approx 0$ which ensures that  $E(\alpha>1/2,h=0)\rightarrow 1$, {\it i.e.}, $p_{\pm}\approx 1/2$; see Fig. 5. It should be remembered that in the ohmic case, the reduction of quantum coherence (superposition) of the spin rather occurs at $\alpha=1/2$ and not at the quantum phase transition. Additionally, the maximal entanglement occurring at $\alpha=1/2$ allows to explain the beginning of the ``incoherent'' dynamical behavior \cite{Leggett}. More precisely, as $\alpha< 1/2$, one has damped Rabi oscillations at low temperatures; the Rabi oscillation frequency $\hat{\Delta}$ and damping rate $\Gamma$ are given by $\hat{\Delta}=\cos\eta T_K(\alpha)$ and $\Gamma=\sin\eta T_K(\alpha)$ with \cite{Leggett} $\eta={\pi \alpha}/{(2(1-\alpha))}$. In contrast, for $1/2<\alpha<1$ the Rabi oscillations disappear and the behavior becomes completely incoherent \cite{Saleur}. When the entanglement with the environment becomes too important  one cannot distinguish in which state, 
$\uparrow$ or $\downarrow$, is the two-level system anymore!

The plateau of entanglement is reminiscent of the maximal entanglement between the spin and its cloud of electrons in the SU(2) Kondo model \cite{Amico,Affleck}.

Second, in the localized phase, we obtain $\langle \sigma_z \rangle \approx -1$ and $\langle \sigma_x \rangle \approx 0$ --- and therefore $E \approx 0$ --- for infinitesimal $h=0^+$.  Since dissipation localizes the spin in the ``down'' state for $h=0^+$, we do not expect the entropy to depend strongly on the external field; the perturbation theory in $\Delta$ predicts $E \sim -(\Delta/\omega_c)^2 \ln (\Delta/\omega_c)$ to leading order for all $h$ \cite{Angela}.  As we approach the phase transition from the localized side, this behavior is replaced by $E \sim -(\Delta/\omega_c) \ln (\Delta/\omega_c)$ in agreement with $\langle \sigma_z\rangle =-1 +{\cal O}(\Delta/\omega_c)$ at the
KT transition. 
This implies a non-universal jump of the entanglement entropy at the KT phase transition. 
This non-universal jump is reminiscent of the non-universal jump in the longitudinal magnetization $\langle \sigma_z\rangle$. 
It should be noted that at finite $h$, the sharp non-analyticity at the quantum phase transition is replaced by a maximum which signifies the crossover regime $h\sim T_K$ \cite{karyn3}. Finally, we like to emphasize that in the delocalized phase, the Kondo scale controls the entanglement between the spin and the bath. 
For $h\ll T_K$, the disentanglement proceeds as $(h/T_K)^2$ whereas for $h\gg T_K$, $E$ vanishes as $(T_K/h)^{2-2\alpha}$, up to logarithmic corrections. 
This
universal behavior stems from $\langle \sigma_z\rangle$ since $\langle \sigma_x\rangle$ evolves very smoothly with the level asymmetry $h$; see Appendix C. Although $E$ is defined at zero temperature, this universality is reminiscent of thermodynamic quantities \cite{Zarand,Weiss}. 

\begin{figure}[!ht]
\begin{center}
\epsfxsize=0.45\textwidth{\epsfbox{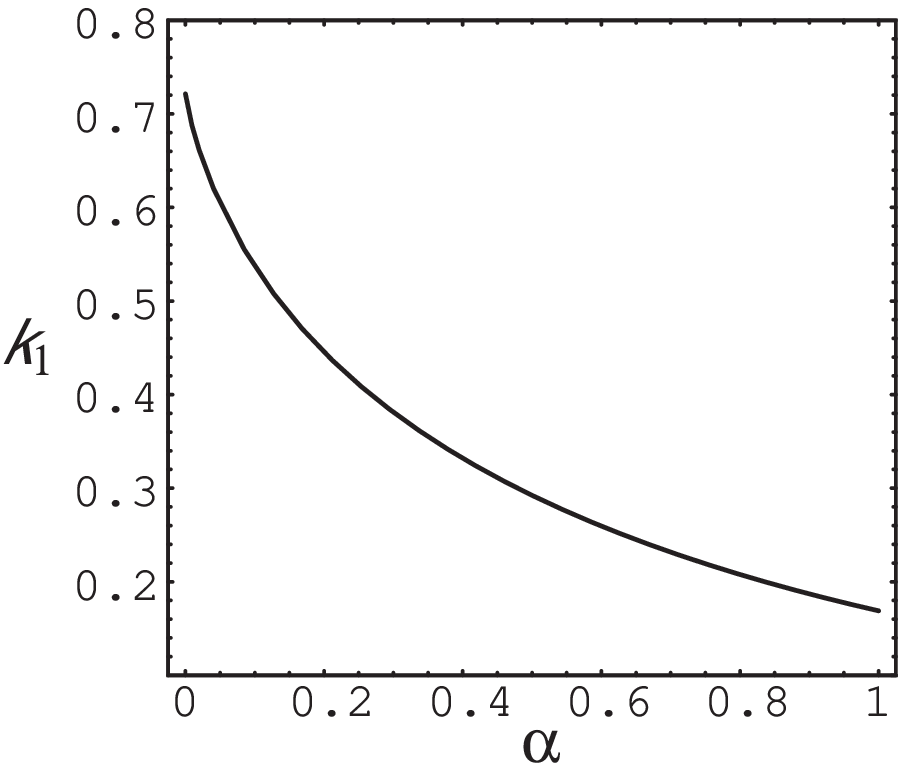}}
\epsfxsize=0.45\textwidth{\epsfbox{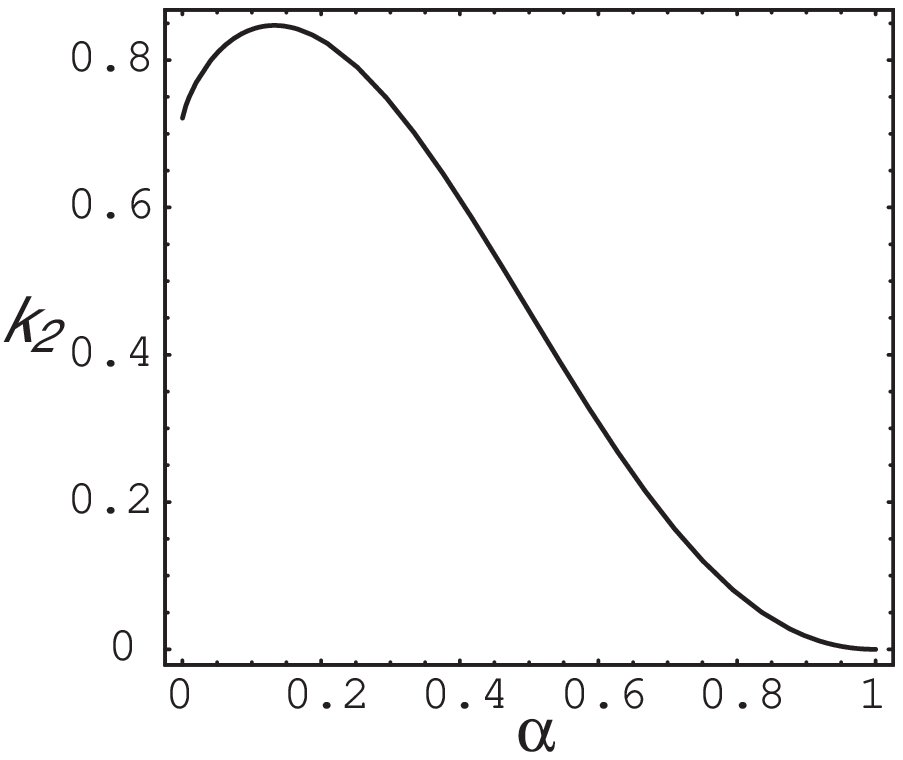}}
\end{center}
\caption{\label{coefficients} Coefficients $k_1$ and $k_2$ as a function of $\alpha$.}
\end{figure}

To be more precise, for a small level asymmetry, we find the scaling,
\begin{equation}
\lim_{h \ll T_K \ll \Delta} E(\alpha,\Delta,h)=E(\alpha,\Delta,0)-k_1(\alpha) \left( \frac{h}{T_K} \right)^2,
\label{Esmh}
\end{equation}
where the coefficient $k_1(\alpha)$ is given by,
\begin{equation}
k_1(\alpha)=\frac{2 e^{\frac{b}{1-\alpha}}}{\pi \ln 2} \left( \frac{\Gamma[1+1/(2-2\alpha)]}{\Gamma[1+\alpha/(2-2\alpha)]} \right)^2,
\end{equation}
and 
\begin{equation}
E(\alpha,\Delta,0)=1-\frac{1}{2 \ln 2} \left( \frac{1}{2\alpha-1} \frac{\Delta}{\omega_c}+C_1(\alpha) \frac{T_K}{\Delta} \right)^2.
\end{equation}
For a large level asymmetry this rather results in 
\begin{equation}
\lim_{T_K \ll h \ll \Delta}E(\alpha,\Delta,h)=k_2(\alpha) \left( \frac{T_K}{h} \right)^{2-2\alpha} \ln \left( \frac{h}{T_K} \right),
\end{equation}
where the pre-factor is given by
\begin{equation}
 k_2(\alpha)=\frac{(1-\alpha)e^{-b}}{\sqrt{\pi} \ln 2} \frac{\Gamma(3/2-\alpha)}{\Gamma(1-\alpha)}.
\end{equation}
The fact that $k_2\rightarrow 0$ when $\alpha\sim 1$ is consistent with the fact that $E$ does not depend much on the level asymmetry in the localized phase. The evolutions of the coefficients $k_1$ and $k_2$ as a function of $\alpha$ are shown in Fig. 6. At the Toulouse point \cite{Toulouse}, one can
easily check that 
\begin{eqnarray}
k_1(\alpha=1/2) = \frac{2}{\pi^2 \ln 2} \\ \nonumber
k_2(\alpha = 1/2) = \frac{1}{\pi \ln 2},
\end{eqnarray}
in agreement with the non-interacting resonant level model of Appendix B.
In Fig. \ref{entropy}, the full solutions of $\langle \sigma_x\rangle$ and $\langle \sigma_z\rangle$ obtained with Bethe Ansatz and with NRG have been used to plot $E$ versus $\alpha$ for different $h$. 

\subsection{Delocalized phase and Boltzmann-Gibbs entropy}

There is a prominent entanglement entropy in the delocalized phase. Even though this von Neumann entanglement entropy is generally not a good measure of entanglement at finite temperature
\cite{Amico}, it is certainly relevant to discuss the evolution of the entropy of the spin at finite temperature.

For temperatures much smaller than $T_K$, the system is almost in a ``pure'' state since all the excited
states only have a small weight $\sim \exp-(T_K/T)$. Thus, one does not expect a dramatic modification in the entanglement entropy of the system for temperatures smaller than $T_K$. This can be seen by considering the entropy of formation of the system \cite{Amico}. Entanglement at finite temperature
has been evidenced in different materials \cite{hit}. On the other hand, if the temperature becomes much larger than the Kondo energy scale, the entropy of the spin should be of purely ``thermal'' origin \cite{Affleck2} and should converge to\footnote{The P(E) theory allows to prove this rigorously. For large temperatures, the (small) coupling $\alpha$ does not affect much the spin observables since the P(E) function converges to that of the uncoupled two-level system (Appendix D). Zero-point fluctuations of the bath become clearly negligible in front of thermal excitations of the bath oscillators.}
$S =-\hbox{Tr}[\rho \log_2 \rho]$ (normalized to 1) where 
\begin{equation}
\rho = \frac{1}{Z_A}e^{-H_A/T};
\end{equation}
here, $H_A$ is the Hamiltonian for the spin alone and $Z_A$ the associated partition function. 
The main coupling with the reservoir becomes of thermal origin. Now, it should be noted that just above $T_K$ the coupling between the bath and the spin produces some corrections in the thermodynamic properties of the interacting resonant level model which produces
a reduction of the thermal entropy of the spin (above $T_K$)  \cite{Zarand,Weiss}:
\begin{equation}
S(T)  -1 \propto -\left(\frac{T_K}{T}\right)^{2-2\alpha}.
\end{equation}
In this sense, the Kondo temperature corresponds to the energy scale at which the thermal entropy of the spin is strongly reduced and is replaced by a prominent entanglement entropy between the spin and its environment. It should be noted that the thermal entropy $S(T)$  exhibits a similar 
power-law behavior as the von Neumann entropy $E(h)$ at zero temperature. 

In the localized phase, there is no entanglement entropy, and thus the spin is only governed by its
thermal Boltzmann-Gibbs entropy.

\section{Subohmic case: Critically entangled system}

Below, we rather focus on the subohmic situation $0<s<1$ which exhibits a second-order quantum phase transition \cite{Bulla,Matthias,karyn4,Subir} by analogy to classical spin chains \cite{Dyson,Kosterlitz} (as a function of $\Delta$). The case $s=1/2$ is of particular interest since it can be realized  through a charge qubit (dot) subject to the electromagnetic noise of an $RLC$ transmission line \cite{Matthias2}; the ohmic case corresponds to an $LC$ transmission line \cite{karyn,Buttiker}. In the limit of small $\Delta$, the $P(E)$ theory of Appendix A suggests that the spin is localized for all $\alpha$.  On the other hand, a quantum critical point $\Delta_c(\alpha)$ \cite{Matthias,karyn4} still exists and a delocalized phase for the spin occurs for $\Delta>\Delta_c$ where $\Delta_c\rightarrow 0$ when $\alpha\rightarrow 0$. It should be mentioned that, at small $s$, some critical exponents are different from those in the classical Ising model \cite{Matthias,MatthiasIsing}.

The order parameter for this delocalized-localized or quantum-classical transition is usually the longitudinal magnetization of the spin $\langle \sigma_z\rangle$ \cite{Matthias} which now vanishes {\it continuously} at the phase transition,
\begin{equation}
\langle \sigma_z\rangle \propto (\Delta_c -\Delta)^{\beta},
\end{equation}
where simple hyperscaling relations easily lead to (see Appendix D):
\begin{equation}
2\beta = \nu(1-s),
\end{equation}
where the correlation length exponent $\nu$ obeys  \cite{Kosterlitz} $1/\nu=\sqrt{2(1-s)}$
for $s\rightarrow 1$ whereas at small $s$, one finds\cite{Matthias} $1/\nu=s$ through a small $s$-expansion. For the subohmic spin-boson model, $\nu \geq 2$, which ensures that $\beta>0$.

To better characterize those second-order quantum phase transitions, we are prompted to examine the entropy of entanglement shared between the spin and its environment. The subohmic model cannot be solved exactly. Then, we will resort to the ``bosonic'' NRG \cite{Matthias,Bulla,Ingersent,karyn2,karyn4} and to hyperscaling relations which are applicable if the fixed point is not trivially Gaussian \cite{Ingersent2}. Technical details are given in Appendix D.  In fact, there is a simple way to conceive that the maximum of $E$ should coincide with the quantum phase transition for the subohmic situation. Starting from the delocalized phase, since the longitudinal magnetization 
$\langle \sigma_z\rangle=0$ at $h=0$,
\begin{equation}
\label{chip}
\frac{\partial E}{\partial \Delta} = -\frac{\chi_{\perp}}{2\ln 2} \ln\left[\frac{1+\langle \sigma_x\rangle}{1-\langle \sigma_x\rangle}\right] <0,
\end{equation}
where we have introduced the transverse susceptibility $\chi_{\perp}=\partial \langle \sigma_x\rangle/\partial \Delta$; this expression is still valid at $\Delta_c$ since for the subohmic case $\langle \sigma_z\rangle$ is zero at the quantum critical point. Since  $\chi_{\perp}=\partial \langle \sigma_x\rangle/\partial \Delta$ and  $\langle \sigma_x\rangle$ are positive quantities, this implies that $\partial E/\partial \Delta <0$ in the delocalized phase. 
\begin{figure}[!ht]
\begin{center}
\epsfxsize=0.65\textwidth{\epsfbox{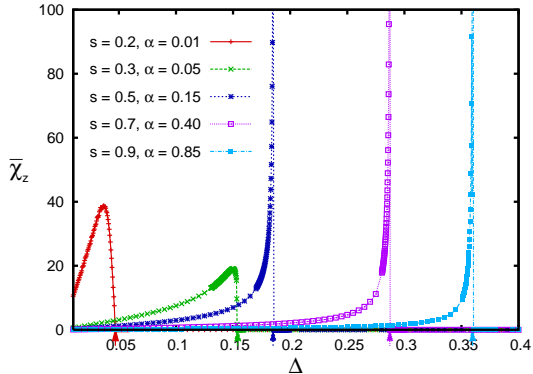}}
\end{center}
\caption{\label{chiy} Susceptibility $\bar{\chi}_z(\Delta)$ in the subohmic spin-boson model.}
\end{figure}
 In the localized phase, $E$ is rather controlled by the finite longitudinal magnetization and by the susceptibility 
$\bar{\chi}_z=-\partial |\langle \sigma_z\rangle|/\partial \Delta>0$ (see Fig. \ref{chiy})\footnote{Here, we neglect the less relevant contribution from $\chi_{\perp}$.}:
\begin{equation}
\label{chiz}
\frac{\partial E}{\partial \Delta} \approx \frac{\bar{\chi}_{z} |\langle \sigma_z\rangle|}{2\ln 2 \langle\sigma_x\rangle} \ln\left[\frac{1+\langle \sigma_x\rangle}{1-\langle \sigma_x\rangle}\right]>0.
\end{equation}
$\bar{\chi}_z$ is related to the rapid localization of the spin. Thus, $\partial E/\partial \Delta >0$ in the localized phase. Eqs. (\ref{chip}) and (\ref{chiz}) imply that the entanglement entropy is maximum at the phase transition. {\it This shows that those second-order impurity quantum phase transitions are always accompanied by a maximum (cusp) in the entropy of entanglement}.
In general, such zero-temperature impurity critical points show a  ``fractional'' entanglement entropy which depends on the dissipation strength $\alpha$ through $\langle \sigma_x\rangle$ and thus is {\it not} universal.  In fact, starting from the localized phase, by analogy to the ohmic case, we find $\langle \sigma_x\rangle = c(\alpha)\Delta/\omega_c$ where $c(\alpha)$ increases by decreasing $\alpha$. 

On the other hand, $E$ unambiguously exhibits universal scalings even though the entanglement is two-sided, so that two numbers are necessary to specify $E$ (one for each of the two ways of approaching $\Delta_c$). More precisely, near the qantum critical point $\Delta_c$, the transverse spin susceptibility obeys 
\begin{equation}
\label{chi1}
\chi_{\perp}(\Delta)= \chi_{\perp}(\Delta_c) +c_{+/-}|\Delta-\Delta_c|^{\zeta},
\end{equation}
\begin{figure}[!ht]
\begin{center}
\epsfxsize=0.65\textwidth{\epsfbox{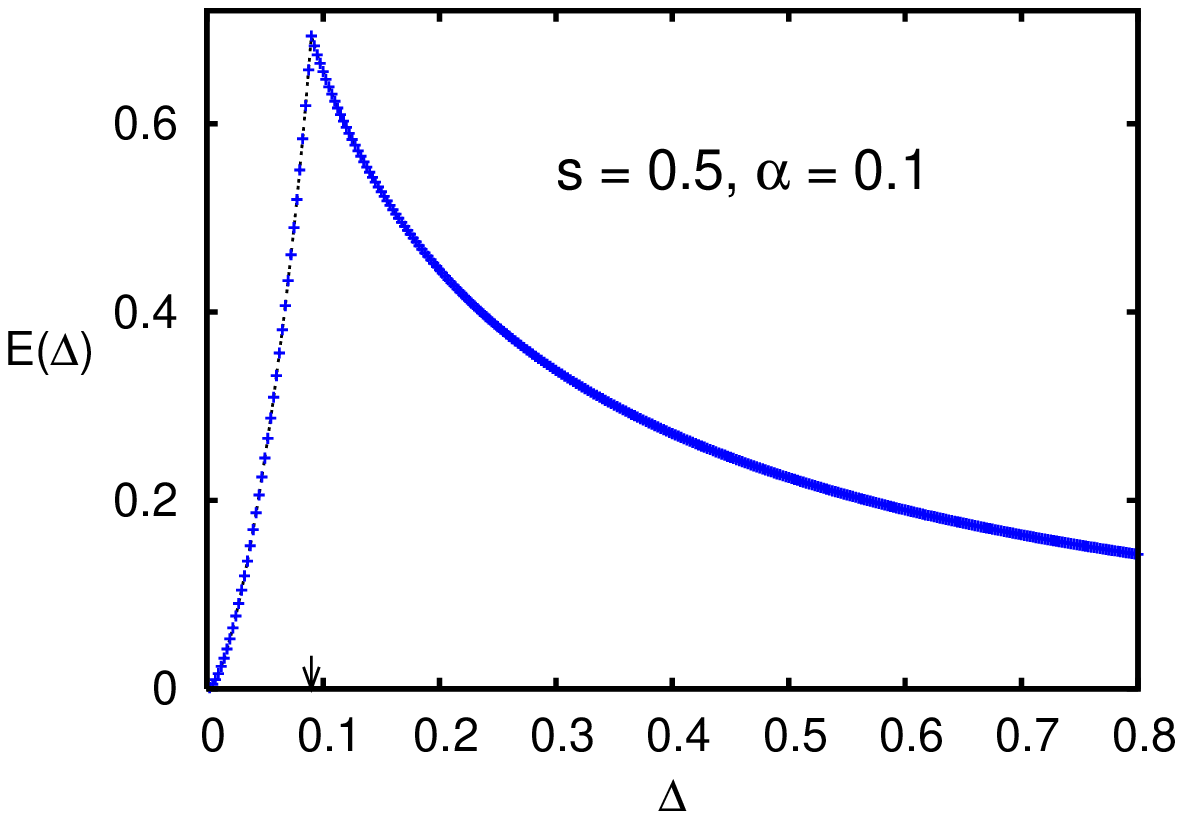}}
\epsfxsize=0.65\textwidth{\epsfbox{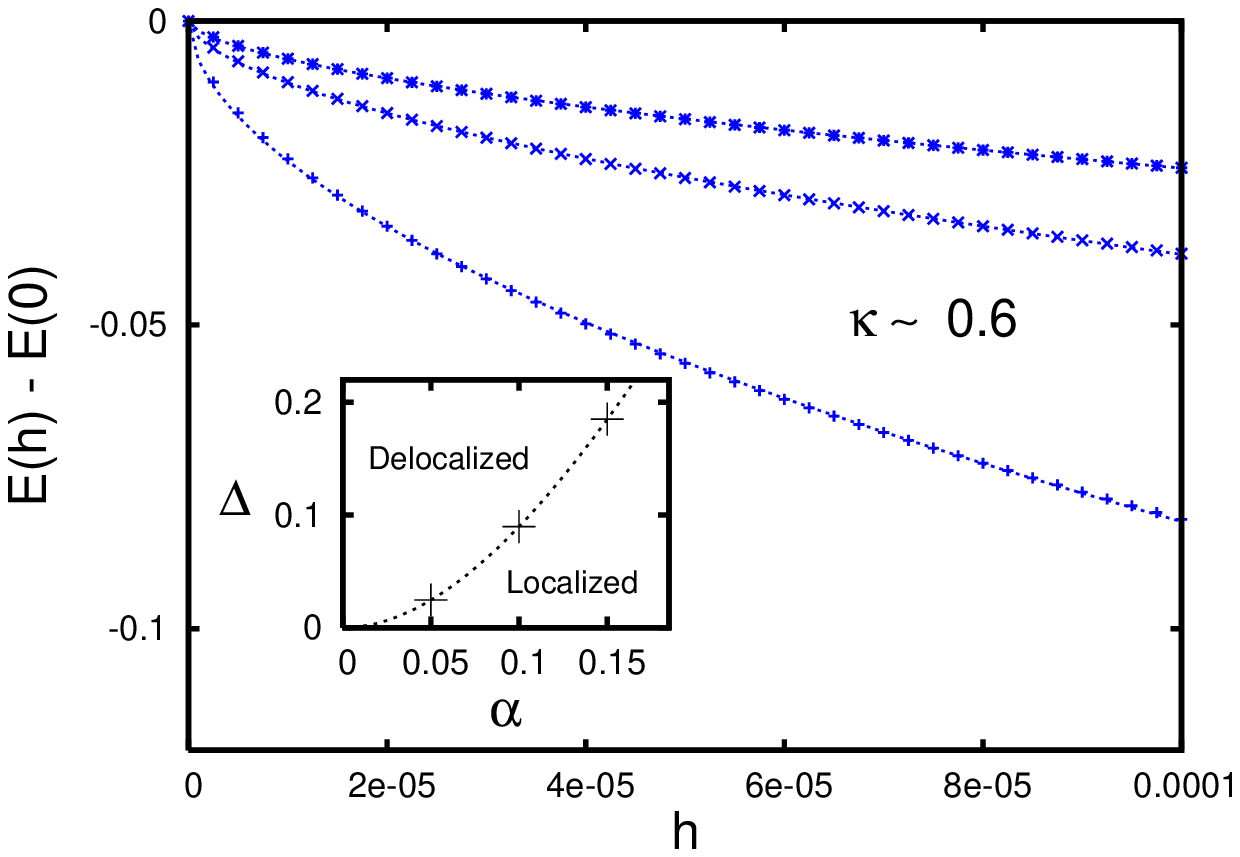}}
\end{center}
\caption{NRG results for the case $s=1/2$, which might be realized through an $RLC$ transmission line. In the subohmic case, $E$ shows a cusp at the phase transition. The NRG reproduces the universal exponent $\kappa=2/3 \sim 0.66$ for different values of $\Delta_c$.} 
\end{figure}
where the exponent $\zeta$ is defined as $\zeta=\nu-2$. For the subohmic spin-boson model,  one finds $\zeta\geq 0$ for all $0<s<1$, ensuring that $\chi_{\perp}$ does not diverge at the transition. Taking into account that $\langle \sigma_x\rangle$ is continuous at the transition, Eq. (\ref{chi1}) thus implies that $E$ always rises linearly for $\Delta\rightarrow \Delta_c^+$
--- $\Delta_c^{\pm}$ means that we approach the quantum critical point from the delocalized (localized) region. It should be noted that the coefficients $c_+$ and $c_-$ can be different in the delocalized and in the localized phase and especially at $s=1/2$ \cite{karyn4}. Through the NRG, we have also checked that $c_+<0$, emphasizing that in the delocalized phase $\chi_{\perp}$ substantially increases at $\Delta_c^+$. For all $0<s<1$, this strongly underlines the duality between the enhancement of entanglement and the strong reduction of the two-spin state quantum superposition near the  phase transition.

In the localized phase, we obtain the scaling behavior:
\begin{equation}
\bar{\chi}_z(\Delta) \propto  |\Delta-\Delta_c|^{-1+\nu(1-s)/2} + a;
\end{equation} 
here $a\neq 0$ when $\nu(1-s)/2>1$, and we identify $a=\bar{\chi}_z(\Delta_c^-)$. For $\Delta>\Delta_c$, $\bar{\chi}_z=0$. Using a small $s$-expansion \cite{Matthias} which predicts $\nu=1/s$ at small $s$,
we deduce that $\bar{\chi}_z$ diverges at $\Delta_c^-$ for $s>1/3$, which is well verified with the NRG approach; see Fig. 7. Using Eqs. (52) and (53), then\footnote{For $s\leq 1/3$, one rather finds $E(\Delta_c)-E(\Delta)\propto |\Delta-\Delta_c|^{\frac{1}{2s}+\frac{1}{2}}$.}
\begin{equation}
E(\Delta_c) - E(\Delta) \propto |\Delta-\Delta_c|^{\nu(1-s)}.
\end{equation}
We observe that the decay of the von Neumann entropy $E$ in the localized phase is faster than linear for all $s>1/2$ and the behavior becomes strictly linear at $s=1/2$. The entanglement entropy at $s=1/2$ is shown in Fig. 8. In the limit $s\rightarrow 1$, we also check that $E$ becomes rapidly suppressed at $\Delta_c^-$ which is a reminiscence of the KT transition (ohmic case, $s=1$).

Now, we shall discuss the scaling of $E$ with the longitudinal field. Using the hyperscaling relations, for all $0<s<1$ we find that (see Appendix D):
\begin{equation}
E(h,\Delta_c)-E(0,\Delta_c) \propto -|h|^{\kappa},
\end{equation}
and 
\begin{equation}
\kappa=\frac{2}{\delta}=2\left(\frac{1-s}{1+s}\right),
\end{equation}
where $\delta$ is defined in a usual way as: $\langle \sigma_z\rangle (\epsilon,\Delta_c) \propto |h|^{1/\delta}$. It is interesting to observe that one can always expand $p_{\pm}(\epsilon)=p_{\pm}(h=0) \pm mh ^{2/\delta}$ at small $h$ and $m>0$ to satisfy $\partial_{h} E(h)<0$ (the field $h$ favors a product state).
It is also relevant to notice that in the delocalized phase,  $E$ decreases as $h^2$ similar to the ohmic case, whereas in the localized phase by approaching the phase transition we rather find a linear decrease of $E$ with $h$.

For the subohmic case, the entanglement entropy allows us to establish important connections between impurity entanglement, quantum decoherence (or strong reduction of the quantum superposition of the two spin states) when approaching the  phase transition from the delocalized phase, and rapid disentanglement in the localized or classical phase for the spin (the spin is rapidly frozen in one classical state due to dissipation). 

{\it We note that the entanglement entropy $E$ of the spin is an interesting new order parameter of those second-order (impurity) quantum phase transitions even though the maximum of entanglement at the quantum critical point is not universal (but scalings are universal)}. 

\section{Comparison with other models}

Here, we compare the entanglement entropy of the subohmic spin-boson model, which yields a parallel with long-range Ising spin chains and exhibits a second-order phase transition, with the entanglement entropy of the quantum Ising model\cite{Osborne} and with the one of the Dicke model \cite{Lambert,Amico}.  

\subsection{One-dimensional Quantum Ising model}

In fact, we argue that the cusp occurring in the subohmic situation of the spin-boson model is very typical of second-order phase transitions and of quantum Ising models. Remember that the spin-boson model
can be mapped onto an effective action where the spin is subject to long-range correlations 
$1/\tau^{1+s}$ in time. Interchanging space and time, one thus expects that the the single-spin (single-site) entanglement properties in the quantum Ising model yields a very similar cusp at the phase transition. 

As an example, let us consider the one-dimensional (1D) quantum Ising model which is described by the following Hamiltonian:
\begin{equation}
H =-\sum_{j=0}^{N-1} \left(\frac{\lambda}{2}\sigma_j^z \sigma_{j+1}^z + \sigma_j^x\right),
\end{equation}
where $\sigma_j^a$ is the $ath$ Pauli matrix (a=x,y or z) at site $j$, and $\lambda$ describes the Ising coupling between spins whereas the transverse field $\Delta$ has been set to $1$. The model can be solved exactly using the Jordan-Wigner transformation \cite{Subir} and the system exhibits a second-order quantum phase transition at $\lambda_c=1$ separating a paramagnetic phase where $\langle \sigma_j^z\rangle =0$ from a ferromagnetic phase. In the ferromagnetic phase, the order parameter obeys (we use the fact that the dynamic critical exponent $z=1$ and $\nu=1$ \cite{Subir}):
\begin{equation}
\langle \sigma_j^z\rangle \propto (\lambda-1)^{1/8}.
\end{equation}
Here, $\beta=1/8$ as a reminiscence of the classical two-dimensional Ising model \cite{Subir}. By analogy to the spin-boson model, one can verify that $\langle \sigma_j^x\rangle$ is {\it continuous} at the second-order quantum
phase transition \cite{Osborne}:
\begin{equation}
\langle \sigma_j^x\rangle =\frac{1}{\pi} \int_0^{\pi} d\phi \frac{1+\lambda\cos\phi}{\sqrt{1+\lambda^2+2\lambda\cos\phi}}.
\end{equation}
In the delocalized phase $(\lambda\ll 1)$, we recover that $\langle \sigma_j^x\rangle \rightarrow 1$ whereas in the localized phase $(\lambda\gg 1)$, we observe that $\langle \sigma_j^x\rangle$ is small
$\sim 1/\lambda$. Similar to the spin-boson model, one can check that the transverse susceptibility $\chi_{\perp} = \partial \langle \sigma_j^x\rangle/\partial \lambda$ does not diverge at the
transition, which implies that in the delocalized phase, the single-spin (single-site) entanglement entropy rises {\it linearly} close to the quantum critical point \cite{Osborne}. In the localized region, from Eq. (55), we also check that $\partial E/\partial \lambda\propto (\lambda-1)^{\frac{2}{8}-1}$ diverges at $\lambda_c$, meaning that $E$ goes very rapidly to zero in the localized phase.  To summarize, the 1D quantum Ising model yields a single-site entanglement entropy which is very similar to that of the subohmic spin-boson model. 

\subsection{Dicke model}

Now, we briefly discuss the entanglement properties \cite{Lambert} in the one-mode superradiance (Dicke) model \cite{Dicke} where collective and coherent behavior of pseudospins (atoms) is induced by coupling --- with interaction $\lambda$ --- to a physically distinct {\it single-boson} subsystem. The Hamiltonian reads \cite{Emary,Dusuel}:
\begin{equation}
H = \omega_o J_z + \omega a^{\dagger} a +\frac{\lambda}{\sqrt{2j}}\left(a+a^{\dagger}\right)(J_+ + J_-),
\end{equation}
where this form follows from the introduction of collective spin operators of length $j=N/2$. in the thermodynamic limit $(N,j)\rightarrow \infty$, the system undergoes a quantum phase transition at a critical coupling $\lambda_c=\sqrt{\omega \omega_o}/2$, at which point the system changes from a large unexcited normal phase to a superradiant one in which both the field and atomic collection acquire
macroscopic occupations \cite{Emary}. In the thermodynamic limit, the problem reduces to a two-mode problem by using the Holstein-Primakoff transformation \cite{Holstein} of the angular momentum operators $J_z=
(b^{\dagger} b -j)$, $J_+=b^{\dagger}\sqrt{2j-b^{\dagger} b}$, and $J_-=J_+^{\dagger}$; here, $b$ and $b^{\dagger}$ are bosonic operators. In the normal phase $(\lambda<\lambda_c)$, expanding the square roots directly this explicitly results in:
\begin{equation}
H^{(n)} = \omega_o b^{\dagger} b - j\omega_o +\omega a^{\dagger}a +\lambda(a+a^{\dagger})(b+b^{\dagger}).
\end{equation}
The problem is solvable going to the position-momentum representation for the two oscillators, $x=(1/\sqrt{2\omega})(a+a^{\dagger})$ and $y=(1/\sqrt{2\omega_o})(b+b^{\dagger})$, with the momenta defined canonically. After diagonalizing the problem, one gets two independent (effective) oscillators, and the energies $\epsilon_{\pm}^{(n)}$
of the two independent oscillator modes read \cite{Emary}:
\begin{equation}
\left(\epsilon_{\pm}^{(n)}\right)^2 = \frac{1}{2}\left(\omega^2+\omega_o^2\pm \sqrt{(\omega_o^2-\omega^2)^2+16\lambda^2\omega\omega_o}\right).
\end{equation}
Crucially, one can see that $\epsilon_-^{(n)}$ remains real only for $\lambda<\lambda_c$; thus 
$H^{(n)}$ remains valid only in the normal phase. In this phase, the ground state energy is given
by $-j\omega_o$ which is $\sim {\cal O}(j)$ whereas the excitation energies $\epsilon_{\pm}^{(n)}$ are
${\cal O}(1)$, implying that the excitation spectrum is quasicontinuous. It should be noted that this normal phase has a definite
parity, since the Hamiltonian commutes with the parity operator $\Pi = \exp(i\pi[a^{\dagger} a +
b^{\dagger} b])$ \cite{Emary}. 

In the superradiant phase $(\lambda>\lambda_c)$, the field and the atomic ensemble acquires macroscopic occupations, thus one has to redefine:
\begin{equation}
a^{\dagger} \rightarrow c^{\dagger}+\sqrt{\alpha},\ b^{\dagger}\rightarrow d^{\dagger}-\sqrt{\beta},
\end{equation}
where $\alpha$ and $\beta$ are ${\cal O}(j)$. One can still diagonalize the problem; the solution
of the two-independent oscillator modes reveal the excitation energies \cite{Emary}:
\begin{equation}
2\left(\epsilon_{\pm}^{(s)}\right)^2 = \frac{\omega_o^2}{\mu^2}+\omega^2\pm\sqrt{\left(\frac{\omega_o^2}{\mu^2}-\omega^2\right)^2+4\omega^2\omega_o^2},
\end{equation}
where
\begin{equation}
\mu = \frac{\omega \omega_o}{4\lambda^2} = \frac{\lambda_c^2}{\lambda^2}.
\end{equation}
Again, one can check that the excitation $\epsilon_{-}^{(s)}$ remains real as long as $\lambda>\lambda_c$, and the ground state energy is given by $-j[(2\lambda^2/\omega)+(\omega_o^2\omega/8\lambda^2)]$. It should be noted that the global symmetry $\Pi$ is broken at the phase transition.

In fact, one may identify $\epsilon_-$ with the excitation energy of a photon branch and $\epsilon_+$
with the excitation energy of an atomic banch: $\epsilon_-$ vanishes at the superradiant transition whereas $\epsilon_+$ remains finite. Moreover, for $\lambda\rightarrow \lambda_c$, from either direction one may identify:
\begin{equation}
\epsilon_-(\lambda\rightarrow \lambda_c) \sim \sqrt{\frac{32\lambda_c^3\omega^2}{16\lambda_c^4+\omega^4}}|\lambda_c-\lambda|^{2\nu},
\end{equation}
where $\nu=1/4$ is the critical exponent describing the divergence of the characteristic length $\xi=\epsilon_-^{-1/2}$. {\it The fact that $\epsilon_-$ vanishes at $\lambda_c$ implies that the Dicke model also exhibits a second-order quantum phase transition.}

In the case of an infinite number of atoms $N\rightarrow \infty$, the entanglement entropy between the atoms and field follows the critical behavior\cite{Lambert} (the reduced density matrix is computed
from the ground state wavefunction):
\begin{equation}
E_{\infty} = -\nu\log_2|\lambda-\lambda_c| = \log_2\xi,\hskip 0.5cm \nu=1/4.
\end{equation}
The entanglement between the atoms and field diverges with the same critical exponent as the characteristic length --- a clear demonstration of critical entanglement. For finite $N$, $E_N$ at the critical point varies as $\log_2 N$ by analogy to conformal field theories in
$(1+1)$ dimensions \cite{c}. 

The normal phase allows a ferromagnetic ordering for the pseudospins $(\langle J_z\rangle\rightarrow -j)$, whereas in the superradiant phase, $\langle J_z\rangle$ decreases progressively and continuously. It might be worth to revisit entanglement properties for an infinite number of oscillators, {\it e.g.}, by investigating entanglement properties of the dissipative quantum Ising chain \cite{MatthiasT,Subir2,PIK}; in this case, the spin order parameter vanishes at the quantum phase
transition. 

\section{Conclusion and Measurement}

A dissipative environment induces a mixed state for the spin (some uncertainty in the spin direction) at zero temperature, which is responsible for quantum decoherence and entanglement entropy. We have shed light on the deep connection between entanglement entropy of the spin with its environment, quantum decoherence, and quantum phase transitions.

\subsection{Brief summary}

An ohmic bosonic environment allows to perform exact calculations by resorting to the Bethe Ansatz solution of the interacting resonant level system \cite{ponomarenko,Level}. Recently, an effort has been
done to generalize the Bethe Ansatz approach for interacting resonant levels out of equilibrium \cite{Natan}. For a weak coupling with the environment, the eigenvalues of the spin reduced density matrix obey $p_+ = 1-{\cal O}(\alpha)$ and
$p_- = {\cal O}(\alpha)$ \cite{Jordan} which ensures that the entanglement entropy of the spin with its environment rises linearly with the coupling $\alpha$. On the other hand, at $\alpha=1/2$, the off-diagonal elements of the spin reduced density matrix become very small $\sim\Delta/\omega_c$ and as a result the system exhibits a plateau at maximal entanglement in the delocalized phase. For the ohmic case, the quantum decoherence of the spin does not occur at the KT phase transition but rather at the Toulouse limit \cite{Toulouse} which also marks the beginning of the incoherent dynamical behavior where Rabi oscillations disappear \cite{Leggett,Saleur}. 
In contrast to the longitudinal spin magnetization which is only controlled by the nature of the Kondo (Fermi-liquid) fixed point, the transverse spin magnetization also depends on irrelevant operators which characterize the localized phase (creation of a single kink-antikink pair in time; see Sec. 4.1). At the KT phase transition, the entanglement entropy exhibits a non-universal jump which is reminiscent of the non-universal jump in the longitudinal spin magnetization $\langle \sigma_z\rangle$ \cite{karyn2,Angela}. In contrast, $\langle \sigma_x\rangle$ is continuous at the 
phase transition. In the localized state, the spin lies in one classical state and almost disentangles from its environment; $p_+\rightarrow 1$ and $p_-\rightarrow 0$. In the delocalized phase, we have also
proven that the Kondo energy scale controls the entanglement properties of the spin-boson model;
for $h\ll T_K$ the entropy decreases as $(h/T_K)^2$ in accordance with the Fermi liquid ground state (or with $\langle \sigma_z\rangle\propto h/T_K$) whereas for $h\gg T_K$, the entanglement goes slowly to zero as $(T_K/h)^{2-2\alpha}$. In fact, the entanglement entropy shows
universal scalings that is reminiscent of thermodynamic quantities \cite{Zarand,Weiss}.

In the subohmic situation, the system exhibits a second-order quantum phase transition (as a function of $\Delta$) and we have shown that the entanglement between the spin and its environment is always enhanced near
the quantum critical point. More precisely, the entanglement entropy of the spin always exhibits a visible
``cusp'' at the quantum phase transition and the quantum decoherence of the spin is progressively destroyed in the vicinity of the phase transition. In this case, we have established clear connections
between impurity entanglement, strong reduction of the quantum superposition of the two spin states
when approaching the phase transition from the delocalized phase, and rapid disentanglement in the
localized (classical) phase for the spin. The entanglement entropy represents an important new order parameter of those second-order quantum phase transitions even though the maximum of entanglement entropy at the quantum critical point is not universal (but, scalings are universal). A similar cusp occurs in the single-site entropy in the quantum Ising model \cite{Osborne,Amico}, emphasing the link between spin-boson and Ising models \cite{Dyson,Kosterlitz}. 

For a finite level asymmetry $h$, the entanglement entropy always yields a maximum characterizing
the departure from the delocalized phase.

Open questions concern the time-evolution of entanglement and the link between entanglement entropy and entropy production out of equilibrium in quantum dot systems \cite{Natan2}. It would also be worth to study the entanglement of a chain of quantum spins with a bosonic bath characterized by a dense spectrum \cite{MatthiasT,Subir2,PIK}, and to compare with the Dicke model where the entanglement between the (pseudo)spins and field diverges at the (second-order) phase transition as $E=\log_2\xi$ where $\xi$ is the characteristic length.

\subsection{Measurement}

The problem of measuring entanglement in many-body quantum systems represents an active field of
research \cite{Gil,Jordan}. An important open question in the study of quantum entanglement is whether it can be measured experimentally.  In fact, the model considered here is realized in noisy charge qubits, composed either of a large metallic dot \cite{karyn} (the single electron box) or a superconducting island \cite{Schon} (the Cooper pair box).  The gate voltage controls the level asymmetry $h$, and $\Delta$ corresponds to the tunneling amplitude between the dot and the lead or the Josephson coupling energy of the junction. If the gate voltage source is placed in series with an external impedance, voltage fluctuations will give rise to dissipation even at zero temperature\cite{karyn}.  An external resistor,
which can be modelled by a long $LC$ transmission line, would describe the ohmic situation $(s=1)$
whereas an $RLC$ transmission line could mimic the subohmic case where $s=1/2$ \cite{Matthias2}.
The parameter $\alpha$ can be varied {\em in situ} when a two-dimensional electron gas acts as the ohmic dissipative environment \cite{Clarke}. In the spin-boson model, $E$ depends only on $\langle \sigma_x \rangle$ and $\langle \sigma_z \rangle$, so it can be constructed from physical observables.  While these quantities would obviously be measured at finite temperature, it is possible to recover the ground state density matrix by extrapolating them to their zero-temperature values.  Charge measurements \cite{Lehnert} yield the quantity $\langle \sigma_z \rangle$, which represents the occupation of the dot or island.  In a ring geometry, the application of a magnetic flux generates a persistent current that is proportional to $\langle \sigma_x \rangle$ \cite{Buttiker}.  Another promising system is the atomic quantum dot, which also permits experimental control of the coupling between the dot and the bosonic reservoir \cite{Recati}.
\\

{\it Acknowledgments:} K.L.H. acknowledges fruitful discussions and collaborations with Ph. 
Doucet-Beaupr\' e, W. Hofstetter, and A. Kopp. K.L.H. is also grateful to M. Vojta, M. B\" uttiker, D.-H. Lee,
and J. Preskill for discussions. This work is supported by NSF through the Yale Center for Quantum Information Physics. 

\appendix{A}
\label{P(E)}
\appendixtitle{Spin-boson model and P(E) theory}

Here, we formulate a precise correspondence between the perturbative calculation at small $\Delta/h\ll 1$ and the well-known $P(E)$ theory of dissipative tunneling problems. From  Sec. 3.1, by definition $P(E)$ obeys:
\begin{equation}
P(E) = \prod_m e^{-|\gamma_m|^2}\sum_{n_{m}} \frac{|\gamma_m|^{2n_m}}{n_m!}\delta\left(E-n_m\hbar\omega_m\right).
\end{equation}
We have explicitly taken into account that $P(E)$ is contributed by all the modes $m$. This expression can be rewritten as
\begin{eqnarray}
P(E) &=& \int \frac{dt}{2\pi} \prod_m \sum_{n_m} \frac{|\gamma_m|^{2n_m}}{n_m!} e^{-|\gamma_m|^2}e^{i\left(E- n_m\omega_m\right)t} \\ \nonumber
&=&  \int \frac{dt}{2\pi} e^{iEt} \prod_m \sum_{n_m}  \frac{{|\gamma_m|}^{2n_m}}{n_m!} e^{-in_m\omega_m t-|\gamma_m|^2} \\ \nonumber
&=& \int \frac{dt}{2\pi} e^{iEt} \prod_m e^{|\gamma_m|^2(e^{-i\omega_m t} - 1)} \\ \nonumber
&=&  \int \frac{dt}{2\pi} e^{iEt}e^{\sum_m |\gamma_m|^2(e^{-i \omega_m t} - 1)}\\ \nonumber
&=& \int \frac{dt}{2\pi} e^{iEt} e^{K(t)},
\end{eqnarray}
where we have defined $K(t)=J(t)-J(0)$ and
\begin{eqnarray}
J(t) &=& \int \frac{d\omega}{\pi} \sum_m \frac{\lambda_m^2}{\omega^2} \delta(\omega-\omega_m) e^{-i\omega t}
\\ \nonumber
&=& 2\int_0^{\omega_c} d\omega \frac{\alpha}{\omega} e^{-i\omega t}.
\end{eqnarray}
At long imaginary times $\tau$, in the ohmic case, this gives rise to the usual function \cite{Nazarov,karyn5}:
\begin{eqnarray}
K(\tau) &=& -2\alpha\int_0^{+\infty} \frac{d\omega}{\omega}(1-e^{-\omega|\tau|})e^{-\omega/\omega_c}
\\ \nonumber
&=& -2\alpha\ln(1+\omega_c|\tau|).
\end{eqnarray}
The logarithm appears due to the summation over the infinite number of modes. At low energy $E$, this
results in $P(E)\propto E^{2\alpha-1}$; there is an orthogonality catastrophe for $\alpha<1$ whereas for
$\alpha>1$,  $P(E)\rightarrow 0$ at $E\rightarrow 0$. {\it $P(E)$ can be interpreted as the probability to ``emit'' the energy $E$ to the bath when flipping the spin}. At finite temperature, one rather gets:
\begin{equation}
K(t) = \alpha \int_{-\infty}^{+\infty} d\omega \frac{J(\omega)}{\omega^2}(n(|\omega|)+\theta(\omega))
(e^{-i\omega t}-1).
\end{equation}
The step-function $\theta(\omega)$ corresponds to the zero-point fluctuations and $n(\omega)$ is
the Bose-Einstein distribution function. If the temperature is large, $K(t)$ is controlled by thermal excitations of the bath oscillators. The real part of $K(t)$ thus becomes $\sim -\alpha Tt$ in the long-time limit. The bath function $P(E)$ becomes a Lorentzian and for
very small $\alpha$ the result converges to that of an uncoupled two-level system.

In the subohmic case, $P(E)$ is more rapidly suppressed at low $E$ which suggests that the spin is always in the localized phase at small $\Delta$.

\appendix{B}
\appendixtitle{Non-interacting resonant level at $\alpha=1/2$}

For $\alpha=1/2$, the spin-boson model is known to be equivalent to a non-interacting resonant
level model. The level Green function obeys:
\begin{equation}
G_d(\omega_n) = \frac{1}{i\omega_n -h +i T_K sgn(\omega_n)},
\end{equation}
and we have introduced the Kondo energy scale at $\alpha=1/2$,
\begin{equation}
T_K = \frac{\Delta^2}{D}.
\end{equation}
The density of states of the impurity thus takes the form:
 \begin{equation}
 n_d(\omega) = \frac{1}{\pi}\Im m G_d(\omega) = -\frac{1}{\pi}\frac{T_K sgn(\omega)}{(\omega-h)^2 +T_K^2}.
 \end{equation}
The associated ground-state energy thus reads:
 \begin{equation}
{\cal E}_g =   \int_{-D}^0 d\omega \omega n_d(\omega) = \int_{-D}^{0} \frac{d\omega}{\pi} \frac{T_K \omega}{(\omega-h)^2 +T_K^2} -\frac{h}{2}.
 \end{equation}
The cutoff $D$ is important to regularize the integral. Now, we can perform the
change of variable $\omega\rightarrow \omega+h$ which results in:
\begin{equation}
{\cal E}_g = \int_{-D}^{-h} \frac{d\omega}{2\pi} \frac{2 T_K \omega}{\omega^2 +T_K^2} 
+h  \int_{-D}^{-h} \frac{d\omega}{\pi} \frac{T_K}{\omega^2 + T_K^2}  -\frac{h}{2}.
 \end{equation}
 Finally,
 \begin{equation}
{\cal E}_g = \frac{T_K}{2\pi}\ln\left(\frac{T_K^2+h^2}{T_K^2 +D^2}\right) +\frac{h}{\pi}\tan^{-1}\left(\frac{-h}{T_K}\right).
 \end{equation}
 Thus, this results in:
\begin{eqnarray}
\langle\sigma_z\rangle 
&=& \frac{-2}{\pi}\tan^{-1}\left(\frac{\Delta^2 D^{-1}}{-h}\right) - sgn(h) \\ \nonumber
&=& -\frac{2}{\pi} \tan^{-1}\frac{h}{T_K}.
\end{eqnarray}
In a similar way, we identify
\begin{equation}
 \langle \sigma_x\rangle\  =  -\frac{2}{\pi}\sqrt{\frac{T_K}{D}} \ln\left(\frac{T_K^2+h^2}{D^2}\right) -\frac{4}{\pi}\sqrt{\frac{T_K}{D}}.
 \end{equation}
 It should be noted that at $\alpha=1/2$ results from perturbation theory agree with the non-interacting
 resonant level picture if we equate $D(\alpha=1/2)=4\omega_c/\pi$; this equality will be proven in Appendix C.
 
First, in the limit $h \ll T_K$, we find 
\begin{eqnarray}
\langle \sigma_z \rangle &\to& -(2/\pi)(h/T_K) \\ \nonumber
\langle \sigma_x \rangle &\to& -(4/\pi)\sqrt{T_K/D}[1+\ln (T_K/D)].  
\end{eqnarray}
The result for $\langle \sigma_z \rangle$ is consistent with the Kondo ground state, where $\mathbf{S}$ is fully screened and $\langle S_z \rangle \propto h/T_K$ at small $h$.  Since both $\langle \sigma_x \rangle$ and $\langle \sigma_z \rangle$ are small, the system is close to maximal entanglement, and assuming $T_K\ll D$, we find the following scaling behavior of the entanglement entropy:
\begin{equation}
\hskip -0.04cm \lim_{h \ll T_K} E(1/2,\Delta,h)=E(1/2,\Delta,0)-\frac{2}{\pi^2 \ln 2} 
\left( \frac{h}{T_K} \right)^2,
\end{equation}
where 
\begin{equation}
E(1/2,\Delta,0)=1-\frac{8}{\pi^2 \ln 2}\frac{T_K}{D}\left[ 1+\ln\left(\frac{T_K}{D}\right)\right]^2.
\end{equation}
The second term in Eq.~(B.10) is a universal function of $h/T_K$, with a quadratic dependence on energy that arises from the Kondo Fermi liquid behavior of $\langle \sigma_z \rangle$. In the opposite limit $h \gg T_K$, we find that 
\begin{eqnarray}
\langle \sigma_z \rangle &\to& -1+2T_K/(\pi h) \\ \nonumber
\langle \sigma_x \rangle &\to& -(4/\pi) \sqrt{T_K/D} \ln (h/D).  
\end{eqnarray}
Again, it is instructive to observe that the leading $h$-dependence of $E$ has a universal form dictated by $\langle \sigma_z \rangle$:
\begin{equation}
\lim_{h \gg T_K}E(1/2,\Delta,h)=\frac{1}{\pi \ln 2} \left( \frac{T_K}{h} \right) \ln \left( \frac{h}{T_K} \right).
\end{equation}

The two scaling functions obtained for $h\ll T_K$ and for $h\gg T_K$ imply that the Kondo energy
scale $T_K$ controls the entanglement between the spin and its environment. On the other hand, because of the logarithmic correction, $E$ approaches zero slowly at large $h$. 

\appendix{C}
\appendixtitle{Bethe Ansatz approach}

Here, we use the Bethe-Ansatz solution of Ponomarenko \cite{ponomarenko} on the interacting resonant level model which was inspired by the resonant-level model constructed by Filyov and Wiegmann \cite{Level} (even though a difference emerges in the number of channels and in the method of $\delta$-function regularization).
In fact, $\langle \sigma_z\rangle$ has been calculated by Ponomarenko:
\begin{eqnarray}
\langle \sigma_z\rangle 
=
\begin{array}{rcl}
-\frac{2}{\sqrt{\pi}}\sum_{n=0}^{\infty} \frac{(-1)^n}{(2n+1)n!}e^{\frac{(2n+1)b}{2(1-\alpha)}}\frac{\Gamma
[1+(2n+1)/(2-2\alpha)]}{\Gamma[1+\alpha(2n+1)/(2-2\alpha)]}\left(\frac{h}{T_K}\right)^{2n+1}, h <   T_K 
&& \\ 
-1 -\frac{1}{\sqrt{\pi}}\sum_{n=1}^{\infty} \frac{(-1)^n}{n!} e^{-nb} \frac{\Gamma[1/2 +(1-\alpha)n]}
{\Gamma[1-\alpha n]}\left(\frac{h}{T_K}\right)^{-2n(1-\alpha)}, h  > T_K, &&
\end{array}
\end{eqnarray}
with the parameter $b=\alpha\ln\alpha +(1-\alpha)\ln(1-\alpha)$. In the limiting case $h\ll T_K$, this results in (the term with $n=0$ dominates):
\begin{equation}
\lim_{h \ll T_K} \langle \sigma_z \rangle = -\frac{2e^{\frac{b}{2(1-\alpha)}}}{\sqrt{\pi}}  
\frac{\Gamma[1+1/(2-2\alpha)]}{\Gamma[1+\alpha/(2-2\alpha)]} \left( \frac{h}{T_K} \right).
\end{equation}
Those expressions are consistent with those at $\alpha=1/2$: $\langle \sigma_z\rangle \rightarrow -(2/\pi)(h/T_K)$ for $h\ll T_K$ and $\langle \sigma_z\rangle \rightarrow -1+2T_K/(\pi h)$ at $h\gg T_K$.

$\langle \sigma_x\rangle$ can be obtained by differentiating the ground state energy ${\cal E}_g$ with
respect to $\Delta$. Since $\langle \sigma_z\rangle = 2\partial {\cal E}_g/\partial h$, the ground state
energy is given by
\begin{eqnarray}
{\cal E}_g
=
\begin{array}{rcl}
&& E_0 +E_1(h), h < T_K \\
&& E_0 + E_1(h=T_K) +E_2(h), h > T_K, 
\end{array}
\end{eqnarray}
where $E_1(h)$ and $E_2(h)$ are obtained by integrating the appropriate pieces of $\langle \sigma_z\rangle$:
\begin{equation}
E_1(h) = \frac{1}{2}\int_0^h dh' \langle \sigma_z\rangle_{h'<T_K}, \ E_2(h) = \frac{1}{2}\int_{T_K}^h dh' \langle \sigma_z\rangle_{h'>T_K}.
\end{equation}
We can determine the integration constant $E_0$ by comparing the ground state energy to the perturbative result at $h\gg T_K$. 

First, we need to evaluate the functions $E_1(h)$ and $E_2(h)$. We obtain:
\begin{eqnarray}
E_1(h) = -\frac{1}{\sqrt{\pi}}&\sum_{n=0}^{\infty}& \frac{(-1)^n}{(2n+2)(2n+1)n!}e^{\frac{(2n+1)b}{2(1-\alpha)}}
\\ \nonumber
&&\times\frac{\Gamma[1+(2n+1)/(2-2\alpha)]}{\Gamma[1+\alpha(2n+1)/(2-2\alpha)]}T_K\left(\frac{h}{T_K}\right)^{2n+2}.
\end{eqnarray}
In particular,
\begin{eqnarray}
E_1(h=T_K) = -\frac{1}{\sqrt{\pi}}&\sum_{n=0}^{\infty}& \frac{(-1)^n}{(2n+2)(2n+1)n!}e^{\frac{(2n+1)b}{2(1-\alpha)}} \\ \nonumber
&&\times\frac{\Gamma[1+(2n+1)/(2-2\alpha)]}{\Gamma[1+\alpha(2n+1)/(2-2\alpha)]}T_K.
\end{eqnarray}
The second integral gives
\begin{eqnarray}
E_2(h) = -\frac{h-T_K}{2} -\frac{1}{2\sqrt{\pi}}&\sum_{n=1}^{\infty}& \frac{(-1)^n}{n!}e^{-nb} \frac{\Gamma[1/2+(1-\alpha)n]}{\Gamma[1-\alpha n]} \\ \nonumber
&& \times \frac{T_K}{1-2n(1-\alpha)}\left[\left(\frac{h}{T_K}\right)^{1-2n(1-\alpha)} - 1\right].
\end{eqnarray}
In the limit $h\gg T_K$, equating the coefficients of the $h^{2\alpha-1}$ terms obtained with
perturbation theory and Bethe Ansatz calculations, one identifies
\begin{equation}
-\frac{1}{4}\Gamma(1-2\alpha)\frac{1}{\omega_c^{2\alpha}} = \frac{1}{2\sqrt{\pi}}e^{-b} \frac{\Gamma(3/2-\alpha)}{\Gamma(1-\alpha)} \frac{1}{2\alpha-1}\frac{1}{D^{2\alpha}}.
\end{equation}
This matches Eq. (26) of the paper by Cedraschi and B\" uttiker \cite{Buttiker}. Moreover, the energy $E_0$ is obtained by comparing the ground state energy at large $h$ to the perturbative result.
For $\alpha\neq 0$ we precisely get\footnote{It should be noted that for $\alpha=0$, we can't set $e^{h/\omega_c} =1$ in the expression (\ref{pert}); instead we must 
take $e^{h/\omega_c}=1+h/\omega_c$. This produces a term $\propto (h/\omega_c)^{2\alpha}$
which becomes a constant at $\alpha=0$ and exactly cancels the $\Delta^2$ term in $E_0$.}:
\begin{equation}
E_0 = \frac{\omega_c}{4} \left(\frac{\Delta}{\omega_c}\right)^2\frac{1}{1-2\alpha} - \frac{1}{2\sqrt{\pi}}
\frac{\Gamma[1-1/(2-2\alpha)]}{\Gamma[1-\alpha/(2-2\alpha)]}e^{-\frac{b}{2(1-\alpha)}}T_K.
\end{equation}
The first term is a remnant of the localized phase.

Gathering all these informations, for $h\leq T_K$, we find:
\begin{equation}
\langle \sigma_x \rangle = \frac{1}{2\alpha-1}\frac{\Delta}{\omega_c}+C_1(\alpha)\frac{T_K}{\Delta}+\frac{T_K}{\Delta}f_1\left( \frac{h}{T_K}, \alpha \right),
\label{sxhltk}
\end{equation}
where
 \begin{eqnarray}
C_1(\alpha) & = & \frac{e^{-\frac{b}{2(1-\alpha)}}}{\sqrt{\pi}(1-\alpha)} \frac{\Gamma[1-1/(2-2\alpha)]}{\Gamma[1-\alpha/(2-2\alpha)]} \; , \nonumber \\
f_1(y,\alpha) &=&-\frac{1}{\sqrt{\pi}(1-\alpha)}\sum_{n=0}^{\infty}\frac{(-1)^n e^{\frac{(2n+1)b}{2(1-\alpha)}}}{(n+1)!} \nonumber \\ \nonumber
&&  \hspace*{0.8in}
\times \frac{\Gamma\left(1+\frac{2n+1}{2(1-\alpha)}\right)}{\Gamma\left(1+\frac{\alpha(2n+1)}{2(1-\alpha)}\right)} \left(\frac{h}{T_K}\right)^{2n+2}\; . \nonumber
\end{eqnarray}
This general (and quite complicated) form of $\langle \sigma_x\rangle$ is consistent with the computation of the persistent current by Cedraschi and B\" uttiker \cite{Buttiker}. 

Now, we will seek to simplify this expression assuming that $h\ll T_K$.
$\langle \sigma_x\rangle$ evolves very smoothly $\sim (h/T_K)^2$ with the longitudinal field (level asymmetry). Additionally, it should be noted that in the limit $h\rightarrow 0$\footnote{For $\alpha\rightarrow 0$, the function $C_1\rightarrow 1$, which also ensures that $\langle \sigma_x\rangle \rightarrow 1$ (and as discussed above, the first term in Eq. (\ref{sigmaxBethe}) or in $E_0$ should not be present when
$\alpha=0$).}:
\begin{equation}
\label{sigmaxBethe}
\langle \sigma_x \rangle(\alpha\neq 0) = \frac{1}{2\alpha-1}\frac{\Delta}{\omega_c}+C_1(\alpha)\frac{T_K}{\Delta}.
\end{equation}
This shows that in the delocalized phase, $\langle \sigma_x\rangle$ is not only a universal function depending on the fixed point
properties, {\it i.e.}, on $T_K$, but it still contains the perturbative term in $\Delta/\omega_c$ typical of
the localized phase. In fact, the second term becomes negligible for $\alpha>1/2$ which ensures
that for $\alpha>1/2$:
\begin{equation}
\langle \sigma_x \rangle = \frac{1}{2\alpha-1}\frac{\Delta}{\omega_c}.
\end{equation}
Moreover, the limit $\alpha\rightarrow 1/2$ must be taken carefully: 
$C_1(1/2)=(4/\pi) \Gamma(1-2\alpha) \to 4/(\pi (1-2\alpha))$, so the two terms combine to give 
\begin{equation}
\langle \sigma_x \rangle \to -(4/\pi) \sqrt{T_K/D} \ln (T_K/D), 
\end{equation}
where we have used $D(\alpha=1/2)=4\omega_c/\pi$. One can check that this agrees with the solution at $\alpha=1/2$ to leading order. 

In the opposite limit, $h\geq T_K$, we find
\begin{equation}
\langle \sigma_x \rangle = \frac{1}{2\alpha-1} \frac{\Delta}{\omega_c}+C_2(\alpha)\frac{T_K}{\Delta}+\frac{T_K}{\Delta} f_2 \left( \frac{h}{T_K}, \alpha \right),
\label{sxhgtk}
\end{equation}
where
\begin{eqnarray}
C_2(\alpha) &=&
\frac{1}{\sqrt{\pi}(1-\alpha)}\sum_{n=1}^{\infty}\frac{(-1)^n e^{-nb}}{n!(1-2n(1-\alpha))} 
\ \nonumber \\
&&  \hspace*{0.8in} \times
\frac{\Gamma\left(\frac{1}{2}+(1-\alpha)n\right)}{\Gamma(1-\alpha n)} \; , \nonumber \\
f_2(y,\alpha) & = & \frac{1}{\sqrt{\pi}} \sum_{n=1}^{\infty} \frac{(-1)^n e^{-nb}}{n!(1-2n(1-\alpha))} \frac{\Gamma\left(\frac{1}{2}+(1-\alpha)n\right)}{\Gamma(1-\alpha n)} \nonumber \\
& &  \hspace*{0.8in} \times \left( 2ny^{1-2n(1-\alpha)}-\frac{1}{1-\alpha}\right) \; . \nonumber
\end{eqnarray}
For $h \gg T_K$, we can again simplify,
\begin{eqnarray}
\hskip -0.4cm \lim_{h \gg T_K} \langle \sigma_z \rangle & = & -1+\left(\frac{1-2\alpha}{2}\right) C_2(\alpha) \left( \frac{T_K}{h} \right)^{2-2\alpha} \\
\hskip -0.4cm \lim_{h \gg T_K} \langle \sigma_x \rangle (\alpha\neq 1/2) & = & \frac{1}{2 \alpha -1} \frac{\Delta}{\omega_c} + C_2(\alpha) \frac{T_K}{\Delta} \left( \frac{T_K}{h} \right)^{1-2\alpha},
\end{eqnarray}
where 
\begin{equation}
C_2(\alpha)=\frac{2 e^{-b}}{\sqrt{\pi} (1-2\alpha)} \frac{\Gamma(3/2-\alpha)}{\Gamma(1-\alpha)}.
\end{equation}
Using the precious relation between the two high-energy cutoffs $D$ and $\omega_c$, one can check the good agreement with the perturbative results of Sec. 3.1.
When $\alpha \to 1/2$, $\langle \sigma_x \rangle$ contains two additional terms which conspire with the other terms to produce the logarithm obtained by perturbation theory; we do not write them because they cancel each other for $\alpha \neq 1/2$.  

\appendix{D}
\appendixtitle{Subohmic case and Second-order phase transitions}

For a second-order impurity quantum phase transition we can apply the following scaling ansatz for the impurity part of the free energy,
\begin{equation}
\label{free}
F_{imp} = Tf(|\Delta-\Delta_c|/T^{1/\nu}, h T^{-b}),
\end{equation}
to relate the critical exponents associated with $E$ to other critical
exponents such as the correlation length exponent $\nu$. Even though the temperature $T$ is introduced for the scaling analysis, the entanglement entropy $E$ (which is defined for a pure state) is evaluated at zero temperature. The crossover from the quantum critical regime to one or other of the stable regimes, defines the energy scale $h^*$ that vanishes at $\Delta_c$ as:
\begin{equation}
h^*\propto |\Delta_c-\Delta|^{b\nu}, 
\end{equation}
It should be noted that the ansatz (\ref{free}) is usually well justified when the fixed point is interacting\cite{Matthias,Ingersent2,karyn4}; for a Gaussian fixed point the scaling function would also depend upon dangerously irrelevant variables. 

Integrating out the boson degrees of freedom induces a long-range interaction in time which results in the following term in the action \cite{Matthias}, ${\cal S}_{int} = \int d\tau d\tau' \sigma_z(\tau) g(\tau-\tau') \sigma_z(\tau')$, where $g(\tau)\propto 1/\tau^{1+s}$ at long times. Since the action is dimensionless, this implies that $\sigma_z$ behaves as $\sim \tau^{(s-1)/2}$. Moreover, from the term which contains
the level asymmetry we also infer that $h$ behaves as $\sim \tau^{(-s-1)/2}$ or $T^{(1+s)/2}$. Thus, this results in:
\begin{equation}
b = \frac{1+s}{2}.
\end{equation}
Defining the exponent $\delta$ as $\langle \sigma_z\rangle (h,\Delta_c) \propto |h|^{1/\delta}$, one also finds:
\begin{equation}
\delta = \frac{1+s}{1-s}.
\end{equation}
This is consistent with the NRG results which predict that the
local susceptibility $\chi_z=\partial |\langle \sigma_z\rangle|/\partial h$ at the quantum critical point diverges as $T^{-s}$. In the small
$s$ limit, this leads to $1/\delta \approx 1-2s+{\cal O}(s^2)$; this can also be recovered by resorting to a small $s$ expansion \cite{Matthias}. This shows that the transverse field term is not important to characterize critical exponents.

Another important exponent is $\beta$ which (here) is defined as
$\langle \sigma_z\rangle \propto (\Delta_c -\Delta)^{\beta}$ in the localized phase. Using the analysis above this results in:
\begin{equation}
\beta = \nu b \delta^{-1} = \nu\left(\frac{1-s}{2}\right).
\end{equation}
$\langle \sigma_x\rangle=-2\partial F/\partial \Delta$ obeys $\langle \sigma_x \rangle (h,\Delta_c) - \langle \sigma_x \rangle (0,\Delta_c)\propto -|h|^{1/\bar{\delta}}$, 
with 
\begin{equation}
1/\bar{\delta}=\frac{2}{1+s}(1-1/\nu).
\end{equation}
In fact, since $1/\bar{\delta}(s)\geq 2/\delta(s)$, we can check that the scaling of the entanglement entropy with $h$ mainly stems from the behavior of the longitudinal spin magnetization. On the other
hand, it is interesting to observe that, since at small $s$ the critical exponent $\nu$ obeys $\nu=1/s$, one also gets $1/\bar{\delta}=2/\delta=\kappa$, which is verified through the NRG for $0<s\leq 1/2$ \cite{karyn4}. 

All critical exponents obey hyperscaling which has been verified by NRG \cite{Matthias,karyn4,Bulla,Ingersent}. We can thus safely conclude that the critical fixed point is interacting for all $0<s<1$. Moreover, the exponents $\beta$ and $\delta$ are distinct from the
ones of the long-range classical one-dimensional Ising model (in coordinate space) with $1/r^{1+s}$ interaction for $s<1/2$ which displays mean-field behavior there.  The naive quantum-classical mapping fails for the subohmic spin-boson model if interactions are long-range in time \cite{Matthias,MatthiasIsing}.

 Now, let us define the susceptibility 
$\bar{\chi}_z=-\partial |\langle \sigma_z\rangle|/\partial \Delta$ in the localized phase; we then obtain:
\begin{equation}
\bar{\chi}_z(\Delta) \propto  |\Delta-\Delta_c|^{-1+\nu(1-s)/2} + a;
\end{equation} 
here $a\neq 0$ when $\nu(1-s)/2>1$, and we identify $a=\bar{\chi}_z(\Delta_c^-)$. Since $\nu=1/s$ at small $s$, we infer that $\bar{\chi}_z(\Delta)$ should diverge at $\Delta_c$ for $s>1/3$. 
This is well verified through NRG; see Fig. \ref{chiy}. Moreover, near the quantum critical point $\Delta_c$, the transverse spin susceptibility $\chi_{\perp}=2\partial^2 F/(\partial \Delta)^2$ obeys 
\begin{equation}
\label{chi}
\chi_{\perp}(\Delta)= \chi_{\perp}(\Delta_c) +c_{+/-}|\Delta-\Delta_c|^{\zeta},
\end{equation}
and from the free energy defined in Eq. (\ref{free}):
\begin{equation}
\zeta=\nu-2. 
\end{equation}
For the subohmic spin-boson model,  one finds $\nu\geq 2$ for all $0<s<1$, ensuring that $\chi_{\perp}$ does not diverge at the transition. 
Taking into account that $\langle \sigma_x\rangle$ is continuous at the transition, Eq. (\ref{chi}) thus implies that $E$ always rises linearly for $\Delta\rightarrow \Delta_c^+$
($\Delta_c^{\pm}$ means that we approach the quantum critical point from the delocalized (localized) region) \cite{karyn4}.  It should be noted that the coefficients $c_+$ and $c_-$ can be different in the delocalized and in the localized phase. Through the NRG, we also check that $c_+<0$, emphasizing that in the delocalized phase $\chi_{\perp}$ substantially increases at $\Delta_c^+$, and that $\chi_{\perp}$ shows a clear singularity for $2\leq \nu<3$ or $1/3<s\leq 0.94$ \cite{karyn4}.

\end{article}

\end{document}